%% file: main.tex
\documentclass[prb,twocolumn,superscriptaddress,amsmath,amssymb]{revtex4-1}

\usepackage[justification=raggedright,singlelinecheck=false]{caption}
    \usepackage{hyperref}

\usepackage{orcidlink}
\usepackage{graphicx}
\usepackage{subcaption}
\usepackage{dcolumn}
\usepackage{multirow, array}
\usepackage{siunitx}

\usepackage{bm}
\usepackage{cleveref}
\usepackage{color}
\usepackage{tikz,xcolor,hyperref}
\usepackage{nameref}
\usepackage{physics}
\usepackage{braket}
\usepackage{lineno}
\usepackage{makecell}
\usepackage{vwcol}
\usepackage{lipsum}
\usepackage[T1]{fontenc}
\usepackage{booktabs}
\AtBeginDocument{\RenewCommandCopy\qty\SI}

\def\be{\begin{equation}}
\def\ee{\end{equation}}
\def\bea{\begin{eqnarray}}
\def\eea{\end{eqnarray}}

\colorlet{darkgreen}{green!50!black}
\colorlet{brightyellow}{yellow!75!red}
\colorlet{orange}{red!50!yellow}
\colorlet{darkblue}{blue!60!black}
\colorlet{darkred}{red!80!black}
\usepackage{array}
\usepackage{array,etoolbox}
\usepackage{fancyhdr}
\usepackage{slashed}

\newcommand{\bno}{\begin{eqnarray*}}
\newcommand{\eno}{\end{eqnarray*}}

\newcommand{\bl}{\begin{large}}
\newcommand{\el}{\end{large}}
\newcommand{\bla}{\begin{Large}}
\newcommand{\ela}{\end{Large}}

\newcommand{\ede}{{\end{document}}}

\def\be{\begin{equation}}
\def\ee{\end{equation}}
\def\bea{\begin{eqnarray}}
\def\eea{\end{eqnarray}}

\begin{document}

\input{paper_body}

\clearpage
\onecolumngrid
\appendix

\input{supp_body}

\end{document}

%% file: paper_body.tex
\title{Gaussian Expansion Method for few-body states in two-dimensional materials }	

\author{Luiz G. M. Ten\'orio\orcidlink{}}
\email{luiztenorio@fisica.ufmt.br}
\affiliation{Department of Physics, Graduate School of Science, Tohoku University, Sendai 980-8578, Japan}
\affiliation{Instituto Tecnológico de Aeronáutica, 12.228-900, São José dos Campos, SP, Brazil} 

\author{André J. Chaves\orcidlink{}}
\affiliation{Physics Center of Minho and Porto Universities (CF-UM-UP) and Department of Physics, University of Minho, P-4710-057 Braga, Portugal}
\author{Emiko Hiyama\orcidlink{0000-0002-6352-5766}}
\email{hiyama@riken.jp}
\affiliation{Department of Physics, Graduate School of Science, Tohoku University, Sendai 980-8578, Japan}
\affiliation{RIKEN Nishina Center for Accelerator-Based Science, Wako, Saitama 351-0198, Japan}

\author{ Tobias Frederico\orcidlink{0000-0002-5497-5490}
} 
\email{tobias@ita.br}

\affiliation{Instituto Tecnológico de Aeronáutica, 12.228-900, São José dos Campos, SP, Brazil}

\begin{abstract} 

We investigate the properties of trions in transition metal dichalcogenides (TMDCs) monolayers using the Gaussian Expansion Method (GEM) adapted to two-dimensional systems. Excitons and trions in monolayer TMDCs with the chemical composition MX$_2$ in the 2H phase are studied systematically. We computed the associated exciton and trion binding energies. We find in addition to the known $J=0$ trion the existence of a bound state with  orbital angular momentum $J=1$.  The results for $J=0$ are benchmarked against existing calculations from the Stochastic Variational Method (SVM) and Quantum Monte Carlo (QMC). Furthermore, we analyze the trion internal structure and geometry through their probability density distributions, accounting for the effects of different material shows  that GEM — widely used in studies of strongly interacting few-body systems — is well adapted to allow comprehensive and computationally efficient investigations of trions and potentially other weakly bound few-body states in layered materials.  In addition, we systematically 
exploit the effect of strain and dieletric environment  in the $J=1$ trion predictions, illustrated for the MoS$_2$ monolayer example.

\end{abstract}
\date{\today}
\maketitle

\section{Introduction}

\begin{figure}[htpb]
\includegraphics[width=1.0\linewidth]{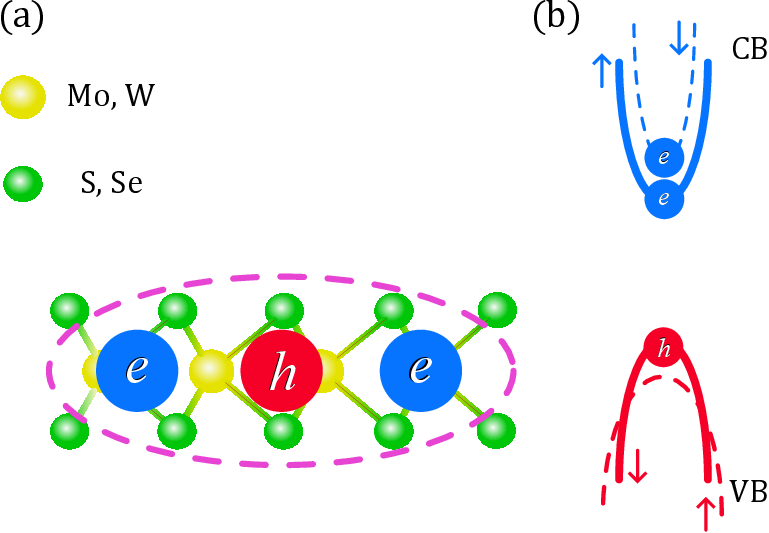}
  \caption{Schematic representation of the trion $X^-$ and TMDCs. In the left panel (a), we illustrate the trion in a single layer TMDC of the form MX$_2$, namely MoS$_2$, MoSe$_2$, Ws$_2$, and WSe$_2$.  In our study  the electrons are in a spin-singlet configuration as presented in the right panel (b), where the dashed lines represent the spin-down band and the full line represents spin-up band, for the conduction band (CB, blue) and valence band (VB, red).}
  \label{fig:Fig_0PRB}
\end{figure}

The emergence of atomically thin semiconductors, particularly monolayer transition metal dichalcogenides (TMDCs), has enabled the observation and control of strongly bound excitonic complexes, which are stabilized by reduced dielectric screening and enhanced Coulomb interactions intrinsic to two-dimensional (2D) systems~\cite{ross2013electrical,chaves2020bandgap,Manzeli2017,lamountain2018environmental}. Among these complexes, trions — charged excitons composed of three charge carriers, see Fig.~\ref{fig:Fig_0PRB} — stand out due to their pivotal role in modifying optical spectra~\cite{opto_1,opto_2} and carrier dynamics in 2D materials~\cite{mak2013tightly,wang2018colloquium}.

Experimental studies have identified negative and positive trions in monolayer TMDCs such as MoS$_2$, WSe$_2$, and MoSe$_2$ through sharp features in photoluminescence and absorption spectra~\cite{ross2013electrical,singh2016trion,courtade2017charged}. Positive (negative) trions are formed in hole (electron) doped TMDCs \cite{courtade2017charged}. They can be inter- or intravalley depending if all the charge carriers are in the same valley or in different ones~\cite{Lyons2019}. Magneto-optics experiments can be used to probe their g-factor, which can reveal information about the valence and conduction band g-factors and many-body effects~\cite{Ashish2021}. These trions exhibit binding energies on the order of tens of meV, which are significantly larger than those in bulk semiconductors~\cite{PhysRevB.13.761,Sergeev2005}. This enhancement originates from the attractive nature of centrifugal potential for zero angular momentum and the effective quasi-2D Coulomb interaction, which is accurately described by the Rytova-Keldysh (RK) potential, incorporating nonlocal screening effects due to the surrounding dielectric environment~\cite{keldysh1979coulomb, Cudazzo}.
Being a composite fermion, the trion has properties suitable for novel technological applications. As neutral particles, excitons have limited control by an external electric field and therefore limited use as an information carrier. The trion propagation, otherwise, can be manipulated at room temperature~\cite{ChoiLeeGong2025}. For TMDCs, trions inherit the valley degree of freedom that characterizes charge carriers in honeycomb lattices, which can be useful in optoelectronic devices and quantum technologies~\cite{Zhang2018,Pham2022,Pu2018}. 

From a theoretical perspective, trions present a genuine three-body problem in an effective quasi-2D Coulomb interaction, demanding careful numerical treatment. Several studies have employed variational approaches, often based on trial wave functions with variational parameters to compute binding energies~\cite{Berkelbach2013}. While providing physical insights, such methods can be limited in capturing the full spatial correlations among particles. Diffusion Monte Carlo (DMC) techniques offer greater accuracy but come at a high computational cost and are less transparent in describing the internal structure of the wave functions~\cite{DMC}. Other popular approaches include a discretized diagonalization in real~\cite{DVR} and momentum-space~\cite{MSRE,FadEq}, which are accurate with guaranteed convergence, however, tend to be very computationally demanding, requiring large physical memory and computational time. In particular, momentum-space methods have difficulties concerning Coulomb interactions and require careful treatment of the numerical convergence~\cite{FadEq}. 

To bridge the gap between accuracy and low-cost, Gaussian-based methods such as the Gaussian Expansion Method (GEM)~\cite{hiyama2003gaussian} and correlated Gaussian basis approaches~\cite{SVM_1,SVM_fewB} have been increasingly applied to few-body systems in condensed matter and nuclear physics. These methods provide a flexible and systematically improvable basis for solving the Schrodinger eigenvalue equation and have shown excellent spectral agreement with both DMC and variational results for excitons and trions in 2D materials~\cite{courtade2017charged}.

In this work, we apply the GEM to investigate the binding energies and spatial properties of trions in monolayer TMDCs. By expanding the three-body wave function over all rearrangement channels in a set of Gaussians with a broad range of widths, we capture both short-range correlations and long-range tail in the presence of nonlocal screening. The RK potential is used to model carrier-carrier interactions, and material-specific parameters, such as effective masses and dielectric constants, are incorporated into a few-body model Hamiltonian. We present results for trion binding energies across different TMDCs and discuss their structure in terms of density functions. Our results are compared against experimental values and prior theoretical models, confirming that GEM offers a computationally efficient and accurate method for exploring trionic effects in 2D semiconductors.

This work is organized as follows. In Sect.~\ref{sec:method}, we detail the Hamiltonian for the  trion in a TMDC layer, the RK potential, as well as the coordinates and conventions used through the work, in addition  the GEM applied to compute the energy eigenvalues and eigenstates are briefly explained by also including the appendix~\ref{app:matelem}, where the detailed form of the matrix elements in 2D  necessary to build the generalized eigenvalue equation that is solved numerically are provided. In Sect.~\ref{sec:trionenergies} the computed trion energies for $J=0$ with GEM for single layer TMDCs: MoS$_2$, MoSe$_2$, WS$_2$ and WSe$_2$ are presented and compared with results from the literature using differcorrent methods. In addition to the $J=0$ energies, we present results for $J=1$ state, and also a detailed discussion of the GEM convergence in one case.  In Sect.~\ref{sec:trionstructure}, we explore the  $J=0$ and $J=1$  trion structure in those suspended single layer TMDCs, presenting results for the density distributions in Jacobi coordinates, different radii and relative angles. 
In Sect.~\ref{sec:strain_dieletric}, we investigate systematically the substrate effects, strain and environmental dielectric
screening,  which are critical for the trions in real TMDC samples.
We close our work in Sect.~\ref{sec:conclusion} presenting a conclusion summary and perspectives for future applications of GEM to compute few-body states in layered TMDCs.

\section{Methodology}\label{sec:method}

For the study of trions in 2D materials, we consider a three-body Hamiltonian for the charge carriers with effective masses and  pairwise interactions.
The interaction includes electron-hole (e-h) attraction and electron-electron (e-e) repulsion, both modeled with the same spatial form $V(r)$ but opposite sign. The Hamiltonian for the negative trion $X^-$ reads:
\begin{multline}\label{eq:eq_ham}
    H =  -\frac{\hbar^2}{2m_e} \nabla^2_{\mathbf{x}_e} 
         - \frac{\hbar^2}{2m_{e}} \nabla^2_{\mathbf{x}_{e'}} 
         - \frac{\hbar^2}{2m_h} \nabla^2_{\mathbf{x}_h}  
       \\ \hspace{-.1cm} +\hspace{-.05cm} V_{eh}\left(\lvert \mathbf{x}_h \hspace{-.05cm}-\hspace{-.05cm} \mathbf{x}_{e'} \rvert\right)   \hspace{-.05cm}+\hspace{-.05cm} V_{eh}\left(\lvert \mathbf{x}_e - \mathbf{x}_h \rvert\right)
         \hspace{-.05cm}+ \hspace{-.05cm}V_{ee}\left(\lvert \mathbf{x}_e - \mathbf{x}_{e'} \rvert\right),
\end{multline}
where $\mathbf{x}_e,\mathbf{x}_e^\prime$ are the electrons coordinates and $\mathbf{x}_h$ the hole coordinate. The effective masses for the electron and hole are $m_e$ and $m_h$, respectively. For the ground-state, we consider the identical particle pair to be in an antisymmetric spin state. We observe that, the Hamiltonian for a positively charged trion, $X^+$, is analogous to Eq.~\eqref{eq:eq_ham} with the proviso that electrons and holes are interchanged. 

The interaction $V_{eh}(r)$ is described by the RK potential~\cite{keldysh1979coulomb}, obtained from the solution of Poisson's equation for a point charge on top of a polarizable sheet and the electron-electron interaction is given by $V_{ee} = -V_{eh}$. The RK potential accounts for nonlocal dielectric screening in 2D materials. The effective interaction captures the deviation from hydrogenic behavior in 2D excitonic systems~\cite{Cudazzo,Nonhydro} and its written as,
\begin{equation}\label{eq:rytovak}
    V_{eh}(r) = -\frac{e^2}{8\epsilon_0\epsilon_m r_0}\left[\mathcal{H}_{0}\left(\frac{r}{r_0}\right)-\mathcal{Y}_{0}\left(\frac{r}{r_0}\right)\right],
\end{equation}
where $e$ is the elementary charge, $\epsilon_0$ denotes the vacuum permittivity, $r_0 = 2\pi\alpha_{2D}/\epsilon_m$ is the screening length, $\alpha_{2D}$ is the in-plane polarizability, and $\epsilon_m = (\epsilon_1 + \epsilon_2)/2$ is the average dielectric constant of the substrate and superstrate. $\mathcal{H}_0$ and $\mathcal{Y}_0$ denote the zeroth-order Struve and Bessel functions of the second kind, respectively. From now on, we consider suspended samples, i.e., $\epsilon_m=1$.

\begin{figure}[htpb]
  \includegraphics[width=1.0\linewidth]{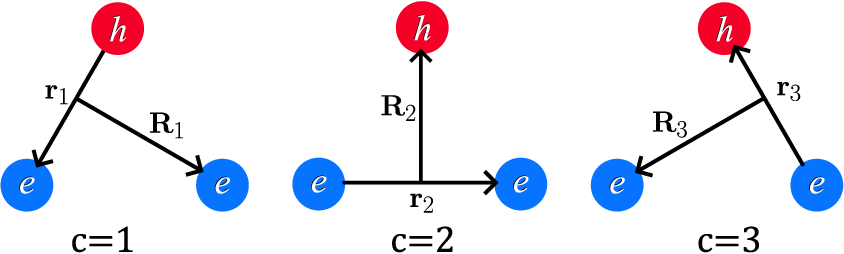}
  \caption{Jacobi coordinates for the three-body system of two electrons and one hole. $\mathbf{r}_c$ denotes the relative coordinates of the two-body subsystem and $\mathbf{R}_c$ labels the spectator particle position relative to the center of mass of the two-body subsystem. The label $c$ indicates the set of Jacobi coordinates. }
  \label{fig:Fig_01PRB}
\end{figure}

The relative Hamiltonian in Jacobi coordinates becomes:
\begin{equation}\label{eq:Red_Ham}
\hspace{-.1cm}    h\hspace{-.05cm} = \hspace{-.05cm}-\frac{\hbar^2}{2\mu_{r_c}}\hspace{-.02cm}\nabla^{2}_{\mathbf{r}_{c}} 
       \hspace{-.02cm} - \hspace{-.02cm}\frac{\hbar^2}{2\mu_{R_c}}\hspace{-.02cm}\nabla^{2}_{\mathbf{R}_{c}}       \hspace{-.025cm}+\hspace{-.025cm} V_{eh}(r_1) \hspace{-.025cm}+ \hspace{-.025cm}V_{eh}(r_3) \hspace{-.025cm}+ \hspace{-.025cm}V_{ee}(r_2),
\end{equation}
where the CM degree of freedom was removed due to translation invariance. The subindex $c$ labels the channel in which the relative kinetic energy operator is constructed.  The Jacobi relative coordinates~\cite{Glckle1983} are given by:
\begin{align}\label{eq:Jacobi}
    \mathbf{r}_{i} &= \mathbf{x}_{k}-\mathbf{x}_{j}, \\
    \mathbf{R}_{i} &= \mathbf{x}_{i} - \frac{m_j \mathbf{x}_j + m_k\mathbf{x}_k}{m_j+m_k},
\end{align}
with
\begin{equation}
\mathbf{X}_{\mathrm{CM}} = \frac{m_i\mathbf{x}_{i} + m_j\mathbf{x}_{j} + m_k\mathbf{x}_{k}}{M},
\end{equation}
and $M = m_i + m_j + m_k$ is the total mass and $i\neq j\neq k$.  The Jacobi relative coordinates are defined cyclically, and the rearrangement channels are illustrated in Fig.~\ref{fig:Fig_01PRB}.

The Hamiltonian is invariant under 2D rotations, making the total orbital angular momentum $J$ a good quantum number. Therefore, we 
 seek for energy eigenstates $\Psi_{J}$ such that:
\begin{equation}\label{eq:SchrodEq}
    h\,\Psi_{J} = -E_{3B}\,\Psi_{J},\quad \widehat{J}_\perp\,\Psi_{J} = J\,\Psi_{J}, \quad J \in \mathbb{Z}.
\end{equation}
where $E_{3B}$ is the trion binding energy and $\widehat{J}_\perp$ is the component of the angular momentum operator transverse  to the plane. 
To each pair of Jacobi coordinates, $\mathbf{r}_i$ and $\mathbf{R}_i$, we associate angular momentum quantum numbers, $\ell$ and $L$ respectively, whose algebra satisfies scalar addition $J = \ell + L$ for angular momentum in two dimensions. 

To solve eigenvalue equations~\eqref{eq:SchrodEq} we adopt the GEM~\cite{hiyama2003gaussian,hiyama_example_1,hiyama_example_2,hiyama_example_3,hiyama_example_4,hiyama_example_5}. In the adopted method, the trion wave function is written as a sum over all rearrangement channels, where for each channel is taken an expansion in a non-orthogonal set of states,
\begin{equation}\label{eq:psitotal}
    \Psi_{J} = \sum_{c,\alpha_c}\mathcal{A}^{c}_{\alpha_c}\ket{\Phi^{c}_{\alpha_c}},
\end{equation}
where the rearrangement channel is denoted by $c$ and the basis functions \{$\ket{\Phi^{c}_{\alpha_c}}$\} labeled by $\alpha_c \equiv \{n_c,N_c,\ell_c,L_c\}$. This is the one of the main differentiating point of GEM from other correlated gaussian methods. The expansion over rearrangement channels allows for a large function space and is more convenient for the case of particles with different masses.
In configuration space  $\braket{\mathbf{r}_c,\mathbf{R}_c|\Phi^{c}_{\alpha_c}}$ is written as:
\begin{small}
\begin{equation}\label{eq:Phic}
   \hspace{-.2cm} \Phi^{c}_{\alpha_c}(\mathbf{r}_c,\mathbf{R}_c) =    \phi_{n_c,\ell_c}(r_c)\,\psi_{N_c,L_c}(R_c)
   \left[Y_{\ell_c}(\hat{\mathbf{r}}_c),Y_{L_c}(\hat{\mathbf{R}}_c)\right]_{J},
\end{equation}
\end{small}
with radial components,
\begin{align}\label{basis_funcs}
    \phi_{n_c,\ell_c}(r_c) &= N_{n_c,\lvert \ell_c \rvert}\,r_c^{\lvert \ell_c \rvert}e^{-\nu_{n_c} r_c^2}, \\
    \psi_{N_c,L_c}(R_c)    &= N_{N_c,\lvert L_c \rvert}\,R_c^{\lvert L_c \rvert}e^{-\lambda_{N_c} R_c^2},
\end{align}
the quantities, $N_{n_c,\lvert \ell_c \rvert}$ and $N_{N_c,\lvert L_c \rvert}$, are normalization factors,
\begin{subequations}
\begin{align} \label{eq:norm-a}
    N_{n_c,\lvert \ell_c\rvert} & = \sqrt{\frac{(4\nu_c)^{\ell_c+1}}{\ell_c!!}}, \\  \label{eq:norm-b}
    N_{M_c,\lvert L_c\rvert} & = \sqrt{\frac{(4\nu_c)^{L_c+1}}{L_c!!}}.
\end{align}   
\end{subequations}
The variational parameters, $\nu_{n_c}$ and $\lambda_{N_c}$, are distributed by a geometric progression labeled by $n_c$ and $N_c$ defined as follows,
\begin{align}\label{var_par1}
    \nu_n = \frac{1}{r_n^2}, \quad
    r_n = r_1 a^n, \\
    \lambda_n = \frac{1}{R_n^2}, \quad
    R_n = R_1 A^n,
\end{align}
where $a$ and $A$ are given by,
\begin{align}\label{var_par2}
a &= \left(\frac{r_{n_{\max}}}{r_1}\right)^{1/(n_{\max}-1)},\\
A &= \left(\frac{R_{n_{\max}}}{R_1}\right)^{1/(N_{\max}-1)}.
\end{align}
The use of the geometric progression for the variational parameters, $r_n$ and $R_N$, yields a compact, densely distributed basis at short distances while also covering the asymptotic region. The combination of compact and the diffuse basis functions allows capturing short-range correlations as well as the long-range tail of the wave function, even in the context of interactions with strong repulsive cores and shallow attractions~\cite{Efimov_GEM,Tetramer_GEM}. The form of the basis function, together with the Infinitesimally Shifted-Gaussian Lobe (ISGL) technique discussed in the supplementary material allows for the analytical treatment of most if not all matrix elements, while also convenient for handling anisotropic mass tensor (for instance, in the case of phosphorene~\cite{Rodin}), which is beyond the scope of the present work.
The angular basis functions are products of spherical harmonics, evaluated in the equatorial plane ($\theta = \pi/2$),
\begin{small}
\begin{equation}
\hspace{-.3cm}    \left[Y_{\ell_c}(\hat{\mathbf{r}}_c),Y_{L_c}(\hat{\mathbf{R}}_c)\right]_{J} 
    = \delta_{\ell_c+L_c,J}\,
      Y_{\lvert \ell_c \rvert,\ell_c}(\hat{\mathbf{r}}_c)\,
      Y_{\lvert L_c \rvert,L_c}(\hat{\mathbf{R}}_c).
\end{equation}
\end{small}

The energy and the coefficients of the eigenstate expansion~\eqref{eq:psitotal} are obtained through the Rayleigh-Ritz variational principle, which leads to a  generalized eigenvalue problem:
\begin{equation}\label{eq_sec}
\sum_{c^{\prime},\alpha_{c^{\prime}}}\left[\mathcal{K}^{(cc^{\prime})}_{\alpha_{c}\alpha_{c^{\prime}}}+\mathcal{V}^{(cc^{\prime})}_{\alpha_{c}\alpha_{c^{\prime}}} + E_{3B}\,\mathcal{N}^{(cc^{\prime})}_{\alpha_{c}\alpha_{c^{\prime}}}\right]\mathcal{A}^{c^{\prime}}_{\alpha_{c^{\prime}}} = 0,
\end{equation}
where the matrix elements $\mathcal{K}^{(cc^{\prime})}_{\alpha_{c}\alpha_{c^{\prime}}}$,$\mathcal{V}^{(cc^{\prime})}_{\alpha_{c}\alpha_{c^{\prime}}}$ and $\mathcal{N}^{(cc^{\prime})}_{\alpha_{c}\alpha_{c^{\prime}}}$ are obtained analytically and given in Appendix~\ref{app:matelem}.

\section{Trion energies}\label{sec:trionenergies}

In this section, we present the results for the trion energies obtained with the use of GEM, focusing first on the convergence of that quantity as additional angular momentum channels are included in the basis functions, and subsequently comparing our calculations with the ones found in the literature.

\begin{table}[ht]
  \centering
  \caption{Variational parameters used for all channels in the GEM.}
  \label{tab:table_metpar}
  \normalsize
  \setlength{\tabcolsep}{8pt}
  \begin{tabular}{@{}l l@{}}
    \toprule
    \textbf{Parameter} & \textbf{Value} \\
    \midrule
    $r_1$ (\AA)             & 0.1 \\
    $r_{n_{\max}}$ (\AA)    & 200 \\
    $R_1$  (\AA)            & 0.1 \\
    $R_{N_{\max}}$  (\AA)   & 200 \\
    $n_{\max}$        & 20 \\
    $N_{\max}$        & 20 \\
    \bottomrule
  \end{tabular}
\end{table}

The secular equation~\eqref{eq_sec} is solved to obtain the $X^-$ states in TMDCs of the form MX$_2$, where M = Mo or W and X = S or Se, as illustrated in Fig.~\ref{fig:Fig_0PRB}(a).
The solution of this generalized eigenvalue problem yields the three-body binding energy $E_{3B}$ and the corresponding eigenstate $\Psi_{J}$. To construct the variational basis, we employ the GEM, using the radial functions defined in Eqs.~\eqref{basis_funcs},  as discussed in the previous section. These functions are parametrized by sets of nonlinear range parameters specified in Eq.~\eqref{var_par1}, determined by geometric progressions as in Eq.~\eqref{var_par2}. The dimension of the total basis is characterized by $n_{\text{max}}$ and $N_{\text{max}}$ functions for the Jacobi coordinates $\mathbf{r}$ and $\mathbf{R}$, respectively, and by range parameters $r_1$, $r_{n_{\text{max}}}$, $R_1$, and $R_{N_{\text{max}}}$. The selection of these range parameters is guided by physical considerations. For systems dominated by long-range interactions, such as trions with Coulomb-like potentials, it is essential to resolve both the singular short-range behavior of the potential, as well as, the extended asymptotic tail of the weakly bound trion. The variational parameters chosen to account for both features are presented in Table~\ref{tab:table_metpar}. 

\begin{table}[ht]
  \centering
  \caption{
    Convergence of three-body energy and binding energy $E_{B} = \lvert E_{3B} - E_{2B}\rvert$, in units of meV, for MoS$_2$ negative trions with $J=0$ and $J=1$ angular momentum computed with 20 basis functions for each Jacobi coordinate.
    The effective masses were $m_e = m_h = 0.5m_0$ and screening length $r_0 = 41.469$\,\AA. The variational parameters used are presented in Table~\ref{tab:table_metpar}.
    The binding energy $E_B$ corresponds to the dissociation threshold into an exciton and a free electron.
  }
  \label{tab:merged_convergence}
  \footnotesize
  \setlength{\tabcolsep}{6pt}
  
  \begin{subtable}[t]{\linewidth}
    \centering
    \caption{$J=0$ state.}
    \begin{tabular}{
      @{} l
      S[table-format=3.3]
      S[table-format=2.3]
      @{}
    }
      \toprule
      {$(\ell, L)$} 
        & {$E_{\mathrm{3B}}$ (meV)} 
        & {$E_B$ (meV)} \\
      \midrule
      (0,  0)   & 588.52 & 33.5 \\
      (1, -1)   & 588.52 & 33.5 \\
      (-1,  1)  & 588.52 & 33.5 \\
      (2, -2)   & 588.66 & 33.6 \\
      (-2,  2)  & 588.71 & 33.7 \\
      (3, -3)   & 588.71 & 33.7 \\
      (-3,  3)  & 588.71 & 33.7 \\
      (4, -4)   & 588.72 & 33.7 \\
      (-4,  4)  & 588.72 & 33.7 \\
      Ref.~\cite{SVM_1} &         & 33.7 \\
      \bottomrule
    \end{tabular}
    \label{tab:table_J0}
  \end{subtable}

  \vspace{1em}

  \begin{subtable}[t]{\linewidth}
    \centering
    \caption{$J=1$ state.}
    \begin{tabular}{
      @{} l
      S[table-format=3.3]
      l
      @{}
    }
      \toprule
      {$(\ell, L)$} 
        & {$E_{\mathrm{3B}}$ (meV)} 
        & {$E_{B}$ (meV)} \\
      \midrule
      (1, 0)    & 515.57 & Not bound \\
      (0, 1)    & 556.11 & 1.08 \\
      (2, -1)   & 556.11 & 1.08 \\
      (-1, 2)   & 556.11 & 1.08 \\
      (3, -2)   & 556.25 & 1.22 \\
      (-2, 3)   & 556.27 & 1.24 \\
      (4, -3)   & 556.27 & 1.24 \\
      (-3, 4)   & 556.27 & 1.24 \\
      \bottomrule
    \end{tabular}
    \label{tab:table_J1}
  \end{subtable}
\end{table}

\subsection{GEM convergence }
We begin by examining the convergence of the ground-state energy with respect to the number of angular momentum configurations, as shown in Table~\ref{tab:merged_convergence} for 20 basis functions for each Jacobi coordinate. The $J = 0$ ground state is constructed using all combinations satisfying $\ell + L = 0$, with angular momentum quantum numbers truncated within the range $-4 \leq \ell, L \leq 4$.
As shown in the table, odd values of $\ell$ and $L$ contribute negligibly to the ground-state energy. This behavior is consistent with the antisymmetric spin configuration of the electron pair, which imposes even-parity symmetry on the spatial wave function and thereby suppresses contributions from odd angular momentum components. Moreover, as expected for Coulomb-bound systems, the dominant contribution arises from the s-wave channel, with higher partial waves contributing progressively less. 

The fast convergence of the binding energy for the  $J=0$ state is a direct consequence of the sum over all rearrangement channels combined with the geometric distribution of nonlinear range parameters, which generates a sufficiently flexible functional space. This allows for an accurate representation of both the long-range tail along $\mathbf{R}_k$ and the short-range two-body correlations in the relative distances $\mathbf{r}_k$. For all TMDCs, the binding energy for the $J=0$ state lies in the interval from 25 meV to 35 meV, dictated by their screening length and effective masses (see Table~\ref{tab:Table_Comp}).

\begin{table*}[ht!]
  \centering
\caption{Comparison of binding energy results for the $J=0$ negative trion ($X^{-}$) obtained using various methods and GEM for common TMDC monolayers: MoS$_2$, MoSe$_2$, WS$_2$, and WSe$_2$. The first column lists the computational methods employed, including the stochastic variational method (SVM)~\cite{SVM_1,SVM_fewB}, variationally optimized orbital approach~\cite{Var_opt}, diffusion Monte Carlo~\cite{DMC}, momentum-space Schrödinger equation (MSRE)~\cite{MSRE}, discrete variable representation (DVR)~\cite{DVR}, hyperspherical harmonics (HH)~\cite{HypSphHar,HypSphHar_2}, path-integral Monte Carlo (PIMC)~\cite{PIMC}, and the Faddeev equations (FE) approach. The second column identifies the TMDC material. The third and fourth columns list the effective electron and hole masses in units of the free electron mass ($m_0$). The fifth column presents the screening length in angstroms. The sixth and seventh columns show the exciton binding energies (in meV) calculated by each method and by our GEM results, respectively. The eighth and ninth columns report the $X^{-}$ trion binding energies (in meV) from each method and from our GEM calculations. The final four rows compare our GEM predictions to experimental values, using parameters taken from the SVM reference~\cite{SVM_1}. All GEM results were obtained using basis configurations with $-4 \leq \ell, L \leq 4$ (see Table~\ref{tab:merged_convergence}), and variational parameters listed in Table~\ref{tab:table_metpar}. }
  \label{tab:Table_Comp}
  \normalsize
  \setlength{\tabcolsep}{6pt}
\begin{tabular}{
  @{}l l
  S[table-format=1.2]
  S[table-format=1.2]
  S[table-format=2.3]
  c c
  c c
  @{}
}
\toprule
\textbf{Method} & \textbf{TMDC}
  & \textbf{\(m_e/m_0\)} & \textbf{\(m_h/m_0\)}
  & \textbf{\(r_0\) (\AA)}
  & \multicolumn{2}{c}{\textbf{\(E_{2B}\) (meV)}} 
  & \multicolumn{2}{c}{\textbf{\(E_B\) (meV)}} \\
\cmidrule(lr){6-7}\cmidrule(lr){8-9}
& & & & & Ref. & GEM & Ref. & GEM \\
\midrule
    \multirow{4}{*}{SVM~\cite{SVM_1}}
      & MoS$_2$  & 0.50 & 0.50 & 41.469 & 555.0 & \textbf{555.03} & 33.7 & \textbf{33.7} \\
      & MoSe$_2$ & 0.54 & 0.54 & 51.710 & 480.4 & \textbf{480.40} & 28.2 & \textbf{28.2} \\
      & WS$_2$   & 0.32 & 0.32 & 37.888 & 523.5 & \textbf{523.53} & 33.8 & \textbf{33.8} \\
      & WSe$_2$  & 0.34 & 0.34 & 45.113 & 470.2 & \textbf{470.17} & 29.5 & \textbf{29.5} \\
    \addlinespace
    \multirow{4}{*}{VOO~\cite{Var_opt}}
      & MoS$_2$  & 0.47 & 0.54 & 44.681 & 526.0 & \textbf{525.98} & 31.6 & \textbf{31.7} \\
      & MoSe$_2$ & 0.55 & 0.59 & 53.162 & 476.7 & \textbf{476.69} & 27.7 & \textbf{27.7} \\
      & WS$_2$   & 0.32 & 0.35 & 40.175 & 508.6 & \textbf{508.55} & 32.4 & \textbf{32.4} \\
      & WSe$_2$  & 0.34 & 0.36 & 47.570 & 456.0 & \textbf{456.02} & 28.3 & \textbf{28.4} \\
    \addlinespace
    \multirow{4}{*}{DMC~\cite{DMC_2}}
      & MoS$_2$  & 0.50 & 0.50 & 41.469& 551.4 & \textbf{555.0} & 33.8$\pm \mathrm{0.15}$ & \textbf{33.7} \\
      & MoSe$_2$ & 0.54 & 0.54 & 51.710& 477.8 & \textbf{480.4} & 28.4$\pm \mathrm{0.15}$ & \textbf{28.2} \\
      & WS$_2$   & 0.32 & 0.32 & 37.888& 519.1 & \textbf{523.5} & 34.0$\pm \mathrm{0.15}$ & \textbf{33.8} \\
      & WSe$_2$  & 0.34 & 0.34 & 45.113& 466.7 & \textbf{470.2} & 29.5$\pm \mathrm{0.15}$ & \textbf{29.5} \\
    \addlinespace

    \multirow{4}{*}{MSRE~\cite{MSRE}}
      & MoS$_2$  & 0.47 & 0.54 & 44.681& 525.98 & \textbf{526.0} & 31.82  & \textbf{31.67} \\
      & MoSe$_2$ & 0.55 & 0.59 & 53.162& 476.69 & \textbf{476.69} & 27.84  & \textbf{27.71} \\
      & WS$_2$   & 0.32 & 0.35 & 40.175& 508.55 & \textbf{508.55} & 32.60  & \textbf{32.42} \\
      & WSe$_2$  & 0.34 & 0.36 & 47.570& 456.02 & \textbf{456.02} & 28.55  & \textbf{28.37} \\
    \addlinespace

    \multirow{4}{*}{DVR~\cite{DVR}}
      & MoS$_2$  & 0.47 & 0.54 & 44.681& 526.0 & \textbf{526.0} & 31.7   & \textbf{31.7} \\
      & MoSe$_2$ & 0.55 & 0.59 & 53.162& 476.7 & \textbf{476.7} & 27.7   &\textbf{27.7} \\
      & WS$_2$   & 0.32 & 0.35 & 40.175& 508.6 & \textbf{508.6} & 34.2 & \textbf{32.4} \\
      & WSe$_2$  & 0.34 & 0.36 & 47.570& 456.0 & \textbf{456.0} & 28.4   & \textbf{28.4} \\
    \addlinespace
    \multirow{4}{*}{HH~\cite{HypSphHar,HypSphHar_2}}
      & MoS$_2$  & 0.35 & 0.43 & 38.620&  & \textbf{544.82} & 32.8 & \textbf{34.4} \\
      & MoSe$_2$ & 0.38 & 0.44 & 51.710&  & \textbf{553.27} & 27.6  & \textbf{27.1} \\
      & WS$_2$   & 0.27 & 0.32 & 37.888&  & \textbf{510.07} & 33.1  & \textbf{33.2} \\
      & WSe$_2$  & 0.29 & 0.34 & 45.113&  & \textbf{459.38} & 28.3  & \textbf{29.1} \\
    \addlinespace
    \multirow{4}{*}{PIMC~\cite{PIMC}}
      & MoS$_2$  & 0.47 & 0.54 & 44.686& 526.5 & \textbf{525.9} & 32.0$\pm \mathrm{0.3}$ & \textbf{31.7} \\
      & MoSe$_2$ & 0.55 & 0.59 & 53.162& 476.9 & \textbf{476.70} & 27.7$\pm \mathrm{0.3}$ & \textbf{27.71} \\
      & WS$_2$   & 0.32 & 0.35 & 40.168& 509.8 & \textbf{508.60} & 33.1$\pm \mathrm{0.3}$ & \textbf{32.43} \\
      & WSe$_2$  & 0.34 & 0.36 & 47.570& 456.4 & \textbf{456.01} & 28.5$\pm \mathrm{0.3}$ & \textbf{28.37} \\
    \addlinespace
    \multirow{1}{*}{FE~\cite{FadEq}}
      & MoS$_2$  & 0.47 & 0.54 &27.05 & 753.0 &  \textbf{757.90}& 49.6 & \textbf{48.30} \\
      \addlinespace
        \hline
          \hline
          \addlinespace
    \multirow{4}{*}{EXP}
      & MoS$_2$  & 0.50 & 0.50 & 41.469 &  &\textbf{555.03}  &32~\cite{Exp_MoS2} & \textbf{33.7} \\
      & MoSe$_2$ & 0.54 & 0.54 & 51.710 &  &\textbf{480.40}  &30~\cite{Exp_MoSe2}& \textbf{28.2} \\
      & WS$_2$   & 0.32 & 0.32 & 37.888 &  & \textbf{523.53} &30~\cite{Exp_WS2} & \textbf{33.8}\\
      & WSe$_2$  & 0.34 & 0.34 & 45.113 &  & \textbf{470.17} &30~\cite{Exp_WSe2} & \textbf{29.5} \\
    \bottomrule
  \end{tabular}
\end{table*}
For the $J=1$ state, we observe similarly fast convergence with respect to the number of angular momentum configurations included in the basis set, as shown in Table~\ref{tab:table_J1}. The bound state is primarily formed by a combination of $(\ell, L) = (1,0)$ and $(0,1)$ channels. Among these, the dominant configuration corresponds to a 1s exciton paired with a p-wave electron, a feature that becomes more evident upon examining the spatial structure of the wave function. This is a weakly bound configuration, with binding energy ranging from 0.39\,meV to 1.44\,meV over all considered TMDCs (see Table~\ref{tab:table_3}). 
The $J=1$ trion wave function in coordinate space is antisymmetric by the exchange of the electrons. In order for the state to be completely antisymmetric we rely on a symmetric spin-valley configuration, which may come into two forms: triplet-triplet ($S=1$,$T=1$) or singlet-singlet ($S=0$,$T=0$). The optical activity of trions is dependent on the excess carrier charges and the optical selection rules. For the direct probing of trions in doped TMD samples, which leads to the formation of attractive and repulsive branches of Fermi-polarons, we need both momentum matching and spin conservation between one of the electrons and holes. For the singlet–singlet case, this condition is naturally fulfilled, as one electron always has the appropriate spin and valley quantum numbers to recombine with the hole, resulting in a finite oscillator strength and an optically bright trion (Fermi-Polaron) state. In contrast, for the triplet–triplet configuration, all possible recombination pathways are blocked either by spin or valley mismatch, or are suppressed by destructive interference between the two equivalent electrons, rendering the corresponding $J=1$ state optically dark within the effective-mass description. However, more advanced techniques, based on nonlinear and/or near-field optics, can probe optically dark states, possibly including the foreseen $J=1$ state~\cite{Exp_opt_1,Exp_opt_2}.

\begin{figure*}[t!]
    \centering
  \centering
  \includegraphics[width=0.8\linewidth]{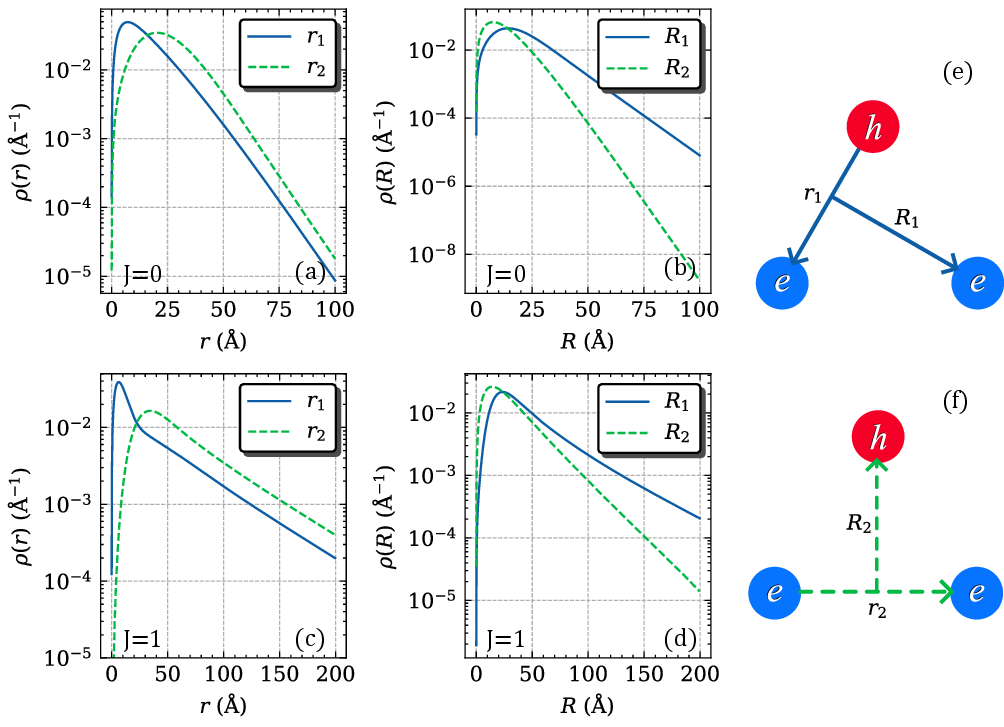}
\caption{Trion $J=0$ and $J=1$ density distributions for a single layer of MoS$_2$  and defined in Eq.~\eqref{eq:densrR}. Top panels (a) and (b) show the densities for $J=0$ state. Bottom panels (c) and (d) show the densities for $J=1$ state.
The solid curves indicate the Jacobi coordinates for  set 1 represented in panel (e) and the  dashed curves label set 2 in panel (f). The parameters used to compute the trion in  MoS$_2$ are $m_e = m_h = 0.5\,m_0$ for the effective masses and $r_0 = 41.469$\,\AA\, for the screening length. }
    \label{fig:4panel}
\end{figure*}

\subsection{Comparison between different methods}

We now compare our results to those obtained with seven different methods in Table~\ref{tab:Table_Comp}. The variational parameters used for GEM are the ones presented in Table~\ref{tab:table_metpar}.  The last column in Table~\ref{tab:Table_Comp} shows our  GEM values for $E_b$, namely the energy separation between the trion $X^-$ and exciton, computed with the effective masses and screening length parameters as reported in each corresponding reference and given in the table. In our calculations, we used 20 basis functions for each Jacobi coordinate and included angular momentum channels up to $|\ell|, |L| \leq 4$, constrained by the condition $\ell + L = J$ as in our previous convergence analysis. 

We start by comparing our results presented in table~\ref{tab:Table_Comp} with the ones obtained by using SVM in Ref.~\cite{SVM_1}. In the SVM approach, the basis for the variational calculation is constructed by using explicitly correlated Gaussians parametrized by a set of nonlinear parameters, which should be computed through a optimization procedure~\cite{SVM_fewB}. The binding energies obtained by GEM show excellent agreement with SVM, alongside with other variational methods such as VOO, where the basis is constructed using a linear combination of Slater-Type Orbitals~\cite{Var_opt}. Both methods differs from GEM, where the basis functions are constructed using Jacobi relative coordinates and the three-body wave function is given by summing over rearrangement channels, while SVM and VOO use single-particle coordinates. Furthermore, as discussed earlier, the nonlinear parameters in GEM are chosen as a geometric progression, they are not computed from an optimization process. The GEM parameters, namely the minimum and maximum range, are chosen to capture the spatial profile of the interactions and at the same time to allow the search for weakly bound states. 

In contrast to the methods discussed above, DVR~\cite{DVR}, and MSRE~\cite{MSRE} formulated in configuration and momentum space, respectively, the Schroedinger eigenvalue equation is discretized over a high-dimensional grid (MSRE) or basis (DVR). The resulting matrix eigenvalue problem is then solved without relying on the variational principle or executing a nonlinear optimization procedure. On the other hand, the FE approach discretize the Faddeev Equations on a large grid and then solves the homogeneous matrix equations by searching for the binding energy. 
The main advantage of these methods is the systematic convergence and ansatz independency, however, the price is paid in computational resources, which grows rapidly with the number of degrees of freedom (eg. Ref.~\cite{MSRE} mentions memory allocation of 190GB with a runtime of 160 hours) turning spatial resolution expensive. Therefore, a systematic agreement of GEM with these methods (see table~\ref{tab:Table_Comp}) provides further insight into the precision of GEM, which is known to be computationally cheap.

Monte Carlo techniques, such as DMC~\cite{DMC,DMC_2} and PIMC~\cite{PIMC}, offer an alternative route to compute trion binding energies. These methods incorporate correlations non-perturbatively but are limited by statistical noise and sampling for weakly bound systems. Our results shown in table~\ref{tab:Table_Comp} align well within the uncertainty bounds quoted in these works, further supporting the reliability of the present application of GEM to three-body systems in 2D materials. In the HH approach~\cite{HypSphHar_2} to the Schroedinger eigenvalue problem, the energy eigenstates are expanded in hyperspherical harmonics. The eigenvalue problem is then reduced to a set of coupled differential equations for the hyperradial function. These equations are coupled through an effective hyperradial potential constructed by projecting the hamiltonian in the hyphespherical harmonic basis~\cite{HypSphHar_2}. The comparison with GEM results for the trion $J=0$ binding energies shows a slight difference for MoS$_2$, MoSe$_2$ and WSe$_2$, while a good agreement for the trion binding energy in the WS$_2$ as presented in table~\ref{tab:Table_Comp}. We call the reader attention that GEM fulfills the variational principle and thus it provides an upper limit to the true trion energy. Therefore, without knowing the exciton energy used to compute the trion binding energy, we cannot judge the small discrepancy between GEM and HH found for the trion in MoSe$_2$.

Finally, comparing with the experimental data~\cite{Exp_MoS2,Exp_MoSe2,Exp_WS2,Exp_WSe2}, we see a reasonable agreement given simplifications pertaining to the effective mass approach to trions. Further developments are necessary in modeling intervalley and intravalley interactions to unravel the trion fine structure associated with the valley degrees of freedom~\cite{Exp_WSe2}.
The inter-method variance of the trion binding energies across all techniques remains within a few meV, comparable to the uncertainty associated with experimental values observed in monolayer TMDCs~\cite{Exp_MoS2,Exp_MoSe2,Exp_WS2,Exp_WSe2}, which are shown in table~\ref{tab:Table_Comp}.

\section{Trion Structure}\label{sec:trionstructure}

Recent advances in experimental techniques have enabled the reconstruction of the intrinsic exciton wave function in momentum space via the momentum distribution of photoexcited electrons~\cite{Man2021}. Through inverse Fourier transform, the corresponding wave function in real space can be obtained. Although this approach has not yet been applied to trions, the same principles could be used to access the internal structure of these states. Motivated by this possibility, we examine in detail the structure of the MoS$_2$ negative trion through the study of its density functions and geometry in configuration space. In addition, we provide the geometrical parameters, radii and relative angles, for the $X^-$ in a single layer of MoSe$_2$,  WS$_2$ and WSe$_2$. In what follows we provide the definitions used for the densities and geometrical  parameters, as well as their values.

\subsection{Density distributions}

The solution of the generalized eigenvalue equation~\eqref{eq_sec} gives the coefficients to reconstruct the terms in the Gaussian expansion the trion wave function in  configuration space~\eqref{eq:psitotal}. Using the coefficients we can calculate analytically the following density distributions as a function of the Jacobi coordinates of a given channel $c$ as follows, 
\begin{subequations}\label{eq:densrR}
\begin{align}
    \rho(r_c) & = 2\pi\, r_c \int_0^\infty\hspace{-.2cm} dR_c \,R_c\int^{2\pi}_0\hspace{-.2cm} d\theta \,\lvert \Psi_J\left(r_c,R_c,\theta\right) \rvert^2,\label{eq:densities_rs} \\ 
    \rho(R_c)& = 2\pi \,R_c \int_0^\infty\hspace{-.2cm} dr_c \,r_c\int^{2\pi}_0\hspace{-.2cm} d\theta\,\lvert \Psi_J\left(r_c,R_c,\theta\right) \rvert^2,\label{eq:densities_rl}
\end{align}\end{subequations}\\
for both $J=0$ and $J=1$ trions.  Eq.~\eqref{eq:densities_rs} corresponds to the density distribution as a function of the relative distance between two particles, namely $e-e$ or $e-h$, while Eq.~\eqref{eq:densities_rl} corresponds to the density distribution as a function of the relative distance between the spectator particle, $e$ or $h$, with respect to the center of mass of the other two.

Figure~\ref{fig:4panel} shows the trion density distributions for both angular momentum $J=0$ and $J=1$ in the MoS$_2$ monolayer.  In panel (a), we present the density distributions for the $J=0$ state as a function of the relative distance between  electron and hole, $r_1$, given by the solid  curve, and the relative distance between the electrons, $r_2$, represented by the dashed curve. As it is expected, the density distribution in the relative distance between the attractive electron and the hole has a more compact profile when compared to the one corresponding to the distribution on the separation of the repulsive electron-electron pair.  As we see, the density distribution as a function  of  $r_2$ is spatially broader with respect to the one in $r_1$,  with a peak around $\sim 20$\,\AA, reflecting the repulsive character of the interaction between the electrons, while in $r_1$ the peak appears around $\sim 5$\,\AA, reflecting the size of the exciton.
The tail of the density distribution as a function of $r_2$  arises from the presence of the hole, which is bound  to the electron-electron pair by anchoring one of the electrons to the excitonic core. We observe that the tail of the density distributions in $r_1$ and $r_2$ present similar slopes at large distances as in both cases the electron is far apart from the $eh$ pair forming the exciton.

Figure~\ref{fig:4panel}(b) displays the density distribution of the spectator particle for the  $J=0$ state. The solid line corresponds to the spectator electron distance to the $eh$ pair, $R_1$, and it shows a broad profile associated to a weak binding of the electron to the exciton with competing attractive and repulsive RK potentials. Conversely, the dashed line, representing the spectator hole distance to the electron-electron pair, exhibits a sharp peak at small $R_2$, indicating that the hole is on average more bound to the electron pair, as both RK potentials to the electrons are attractive. In these cases we observe that the slopes of the tail are different, reflecting the weaker and stronger attraction for the electron and hole as spectators. 

The density distributions for the $J=1$ trion are shown in Fig.~\ref{fig:4panel} as a function of the different Jacobi relative coordinates. We found that the dominant structure of the trion corresponds to a spectator electron in a $L=1$ state weakly bound to the $\ell=0$ exciton, which is expressed in Table~\ref{tab:table_J1} by the leading $(0,1)$ configuration that forms the $J=1$ bound state. The peaks in the density distributions as a function of the relative distance between the $e-h$  and the $e-e$, show the same features as observed for the $J=0$ state, namely their positions are around  $r_1\sim 10$\,\AA and $r_2\sim 40$\,\AA, respectively. The slope of the tail seems to be same, as we have explained for $J=0$ case. 

The spectator electron or hole density distributions for $J=1$ are presented  in Fig.~\ref{fig:4panel}(d), and their respective dependence on $R_1$ and $R_2$, show a prominent peaked structure, qualitatively similar to the ones found for the $J=0$ state but dislocated to larger distances. This is a direct consequence of the dominance of the exciton-like substructure, which governs the dynamics even in this higher angular momentum state and the weaker binding with respect to the $J=0$ case. The tails also have different slopes as already explained for the $J=0$ case.

\begin{figure}[t!]
  \includegraphics[width=8cm]{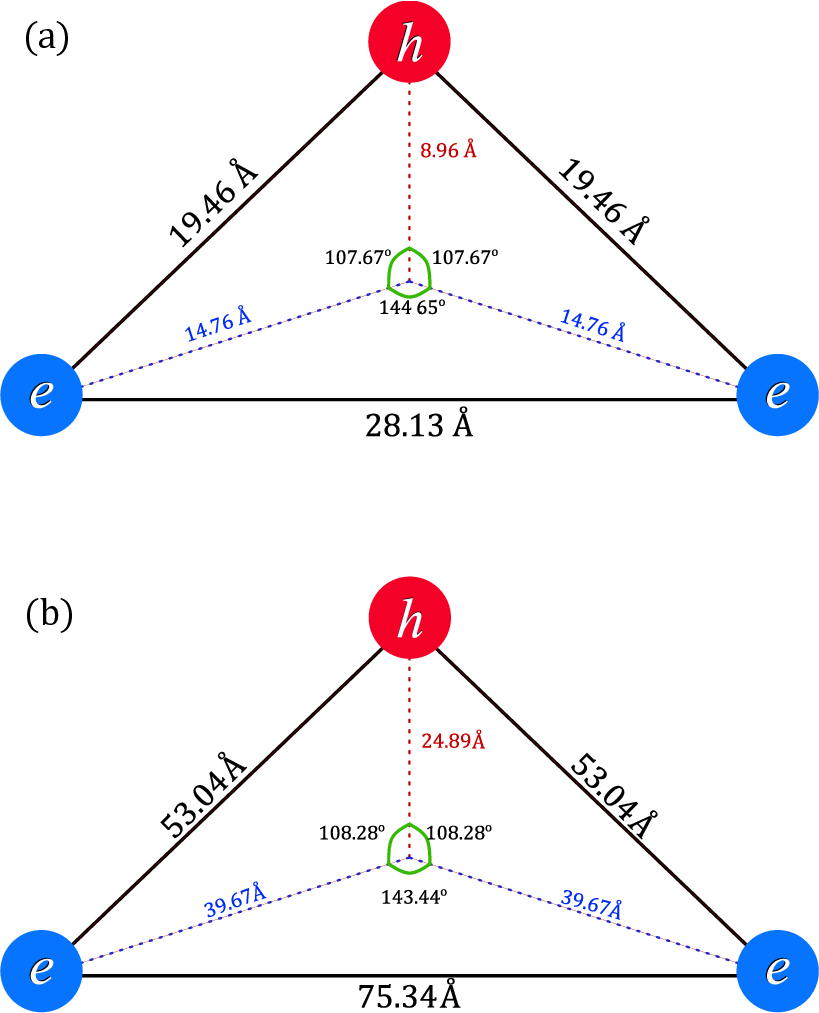}
  \caption{Trion geometry in MoS$_2$. Panel~(a) corresponds to $J=0$ and  panel~(b) to $J=1$ states, respectively. These figures are obtained with $m_e = m_h = 0.5 m_0$ and $r_0 = 41.469$~\AA. }
  \label{fig:geomJ_panels}
\end{figure}

\begin{table*}[t!]
  \centering
  \caption{Trion root-mean-square radii, relative angles and trion-exciton energy separation   in monolayer TMDCs for total angular momentum $J=0$ and $J=1$. The model parameters were taken from the SVM~\cite{SVM_1} calculations, presented in Table~\ref{tab:Table_Comp}.}
  \label{tab:table_3}
  \normalsize
  \setlength{\tabcolsep}{6pt}
  \begin{tabular}{@{} l 
                  *{4}{S[table-format=2.2]}
                  @{\hskip 14pt}
                  *{4}{S[table-format=2.2]} @{}}
    \toprule
    \multirow{2}{*}{Quantity} 
      & \multicolumn{4}{c}{$J = 0$} 
      & \multicolumn{4}{c}{$J = 1$} \\
    \cmidrule(lr){2-5} \cmidrule(lr){6-9}
      & {$\mathrm{MoS}_2$} & {$\mathrm{MoSe}_2$} & {$\mathrm{WS}_2$} & {$\mathrm{WSe}_2$} & {$\mathrm{MoS}_2$} & {$\mathrm{MoSe}_2$} & {$\mathrm{WS}_2$} &{$\mathrm{WSe}_2$} \\
    \midrule
    $\sqrt{\langle r_1^2 \rangle}$ (\AA)    & 19.46 & 20.70 & 23.90 & 25.00 & 53.04 & 52.15 & 80.97 & 75.38 \\
    $\sqrt{\langle r_2^2 \rangle}$ (\AA)    & 28.13 & 30.08 & 34.52 & 36.17 & 75.34 & 74.12 & 114.56 & 106.76\\
    $\sqrt{\langle R_1^2 \rangle}$ (\AA)    & 22.15 & 23.66 & 27.18 & 28.47 & 61.48 & 58.54 & 90.55 & 84.38 \\
    $\sqrt{\langle R_2^2 \rangle}$ (\AA)    & 13.45 & 14.22 & 16.54 & 17.25 & 38.67 & 36.69 & 57.23 & 53.22 \\
    $\theta_{eh}$ ($^{\circ}$)              & 107.67 & 107.49 & 107.71 & 107.64 & 108.33 & 108.26 & 108.42 & 108.38 \\
    $\theta_{ee^{\prime}}$ ($^{\circ}$)     & 144.65 & 145.02 & 144.58 & 144.71 & 143.33 & 143.47 & 143.15 & 143.23 \\
    $\sqrt{\langle \rho_h^2 \rangle}$ (\AA) & 8.96  & 9.48  & 11.02 & 11.50 & 25.78 & 24.46 & 38.15 & 35.47 \\
    $\sqrt{\langle \rho_e^2 \rangle}$ (\AA) & 14.76 & 15.77 & 18.12 & 18.98 & 40.98 & 39.03 & 60.37 & 56.25 \\ 
    $E_B$ (meV)                             & 33.7  & 28.2  & 33.8  & 29.5  & 1.24  & 1.44  & 0.39  & 0.65  \\
    \bottomrule
  \end{tabular}
\end{table*}

\subsection{Trion geometry}

We now explore the geometric structure of the trion. For that aim  the root-mean-square (RMS) values of the Jacobi coordinates, as well as the RMS distances between each particle and the center of mass are calculated. These quantities are defined as,
\begin{align}
\langle \boldsymbol{\rho}_i^2 \rangle 
&= \left\langle \left( \mathbf{x}_i - \mathbf{X}_{\text{CM}} \right)^2 \right\rangle \notag \\
&= \frac{1}{M^2} \bigg[
     m_j^2 \langle r_k^2 \rangle 
   + m_k^2 \langle r_j^2 \rangle \notag \\
&\quad
   + m_j m_k \left( 
         \langle r_j^2 \rangle 
       + \langle r_k^2 \rangle 
       - \langle r_i^2 \rangle 
     \right) 
\bigg]\,.
\label{eq:rho_i}
\end{align}
We remind that the total mass is \( M  \), the center of mass coordinate is  \( \mathbf{X}_{\text{CM}} \) and  each particle coordinate is  \( \mathbf{x}_i \). The expectation values are calculated using the full three-body wave function,
\begin{equation}\label{eq:r2k}
    \langle r^2_k \rangle = \int d\mathbf{r}_k\,d\mathbf{R}_k \, r^2_{k} \, \lvert \Psi_{J} \rvert^2.
\end{equation}

The relative angle between the  center of mass coordinates of two particles, $ \boldsymbol{\rho}_i$ and $\boldsymbol{\rho}_j$, is obtained from the  expectation values of the different radii given in Eq.s~\eqref{eq:rho_i} and \eqref{eq:r2k}:
\begin{equation}\label{eq:ang_ij}
    \theta_{i,j} = \cos^{-1} \left( \frac{ \langle \rho_i^2 \rangle + \langle \rho_j^2 \rangle - \langle r_k^2 \rangle}{ \sqrt{ \langle \rho_i^2 \rangle \langle \rho_j^2 \rangle } } \right)\,.
\end{equation}
The sum over the angles is 180$^\circ$, which also gives a cross-check of the accuracy of our results.

A geometric picture of the trion can be constructed once the lengths of each side of the triangle — with vertices located at the positions of the three carriers — are known. These are given by the RMS values \( \langle r_k^2 \rangle \), defined in Eq.~\eqref{eq:r2k}, and the relative angles \( \theta_{i,j} \), computed using Eq.~\eqref{eq:ang_ij}. Figure~\ref{fig:geomJ_panels} illustrates the geometric configuration of the \( J=0 \) and \( J=1 \) trion states \( X^{-} \) in a MoS$_2$ monolayer, with binding energy shifts of 33.7\,meV and 1.24\,meV relative to the exciton, respectively. One immediately observes that the sum of the internal angles is 180$^\circ$ for both angular momentum states.

It is worth noting that the internal angles are nearly identical for the \( J=0 \) and \( J=1 \) states, suggesting that the weak binding relative to the exciton in both cases places the electron in a region where the effects of the centrifugal potentials are of minor importance, aside from determining the energy separation between the trion and the exciton. The RMS separation between the electrons is approximately 2.5 times larger than the electron–hole separation. This reflects the asymmetric nature of the pairwise interactions: the Coulomb repulsion pushes the electrons apart, while the hole acts as a "glue" binding the three-body system. This picture is further supported by the average position of each particle relative to the center of mass, with the electrons lying about 1.8 times farther away than the hole. It should be emphasized that this is an effective picture, emerging from expectation values over a quantum state with degenerate rearrangement channels due to the indistinguishability and fermionic nature of the electrons. The actual wave function includes a coherent superposition of two indistinguishable configurations, each representing an exciton bound to a spectator electron. The antisymmetrization of the wave function implies that both configurations contribute equally to the state (see, e.g., Ref.~\cite{FadEq}).

We observe that the geometric picture of the trion should be complemented by the density distributions (see Fig.~\ref{fig:4panel}), which express the quantum nature of the bound state as being spread out over a larger spatial region. Therefore, considering that the size of these states is on the order of tens of ångströms, compared to the MoS$_2$ monolayer lattice constant of a few ångströms~\cite{Ataca2012StableSM}, the associated wave function is delocalized over a large spatial region, covering an area two to three orders of magnitude larger than that of the unit cell.

Although our illustrative picture has been restricted to the trion in a MoS$_2$ monolayer, we also compute the geometric elements of the \( J=0 \) and \( J=1 \) trion bound states in single layers of WS$_2$, MoSe$_2$, and WSe$_2$, along with the corresponding energy shifts between the trion and exciton. These results are organized in Table~\ref{tab:table_3}. Owing to the repulsion between the electrons and the attraction within the $eh$ subsystem, we find that \( \theta_{eh} < \theta_{ee} \) is a common characteristic. A striking observation from the table is the near invariance of the internal angles across different materials, despite significant variations in the RMS radii, input charge carrier masses, and RK potential parameters (see the last four rows of Table~\ref{tab:Table_Comp}). The $e$ and $h$ masses in these calculations are equal and the RK potentials for $ee$ and $eh$ just have the same magnitude and opposite sign. We anticipate that perturbing this subtle balance — by violating either of these two conditions — would naturally result in a modification of the internal angles. For instance, in the hypothetical limit where the electron mass is significantly larger than that of the hole, the angle between the electrons tends toward 180$^\circ$, reflecting their symmetric placement on opposite sides of the center of mass.

The properties of the \( J=0 \) and \( J=1 \) trions in monolayers of MoS\(_2\), MoSe\(_2\), WS\(_2\), and WSe\(_2\) are compared in Table~\ref{tab:table_3}.  The exciton-trion energy shift \( E_B \) is approximately 30\,meV for the \( J=0 \) state, while for the \( J=1 \) trion it ranges from 0.39\,meV to 1.44\,meV. The  RMS radii for the \( J=1 \) states are  about 2 to 3 times larger than the ones computed for $J=0$ and approximately inversely proportional to \( \sqrt{E_B} \), as expected from dimensional analysis.

\section{ Strain and dielectric effects }
\label{sec:strain_dieletric}

The model calculations presented so far assumed suspended samples and in this section, we systematically
account for substrate effects, strain and environmental dielectric
screening, all of which are critical in real TMDC samples~\cite{chaves2020bandgap,Strain_tuning,raja2017coulomb}.  Our analysis delineate the limits of the direct applicability of these results to real systems, and in what follows, we explore systematically the 
$J=1$ predictions to incorporate a more realistic modeling of environmental effects.

\begin{table}[ht]
  \caption{Estimative of the strain effect  in the binding energy of the $J=0$ and $J=1$ negative trion states in suspended MoS$_2$ monolayer. We use the parameters from~\cite{Berkelbach2013} for the unstrained layer and an effective mass $(m_e=m_h)$ variation of -4\% for a biaxial 2\% strain $(\tau)$ as indicated by the ab-initio calculations for $m_e$ given in Fig.~4a of Ref.~\cite{Strain_mass}.  The polarizability is then tuned to keep fixed the exciton binding energy with strain~\cite{Strain_Review} to the particular value of $E_{2B}=555.03$\,meV  (see Table~\ref{tab:Table_Comp})  as a function of the effective mass.}
  \label{tab:model_params}
  \normalsize
  \resizebox{\columnwidth}{!}{
  \setlength{\tabcolsep}{6pt}
  \begin{tabular}{@{}c c c c c@{}}
    \toprule
    $\tau$ (\%) & $m$ ($m_0$) & $\alpha$ (\AA) & $E_{J=0}$(meV)& $E_{J=1}$(meV) \\
    \midrule
    0.0\%  & 0.500 & 6.60 & 33.70 & 1.24 \\
    0.5\%  & 0.495 & 6.58 & 33.75 & 1.23 \\
    1.0\%  & 0.490 & 6.55 & 33.80 & 1.21 \\
    1.5\%  & 0.485 & 6.53 & 33.86 & 1.19 \\
    2.0\%  & 0.480 & 6.50 & 33.91 & 1.16 \\
    \bottomrule
  \end{tabular}
  }
\end{table}

\begin{figure}[t!]
  \includegraphics[width=1.0\columnwidth]{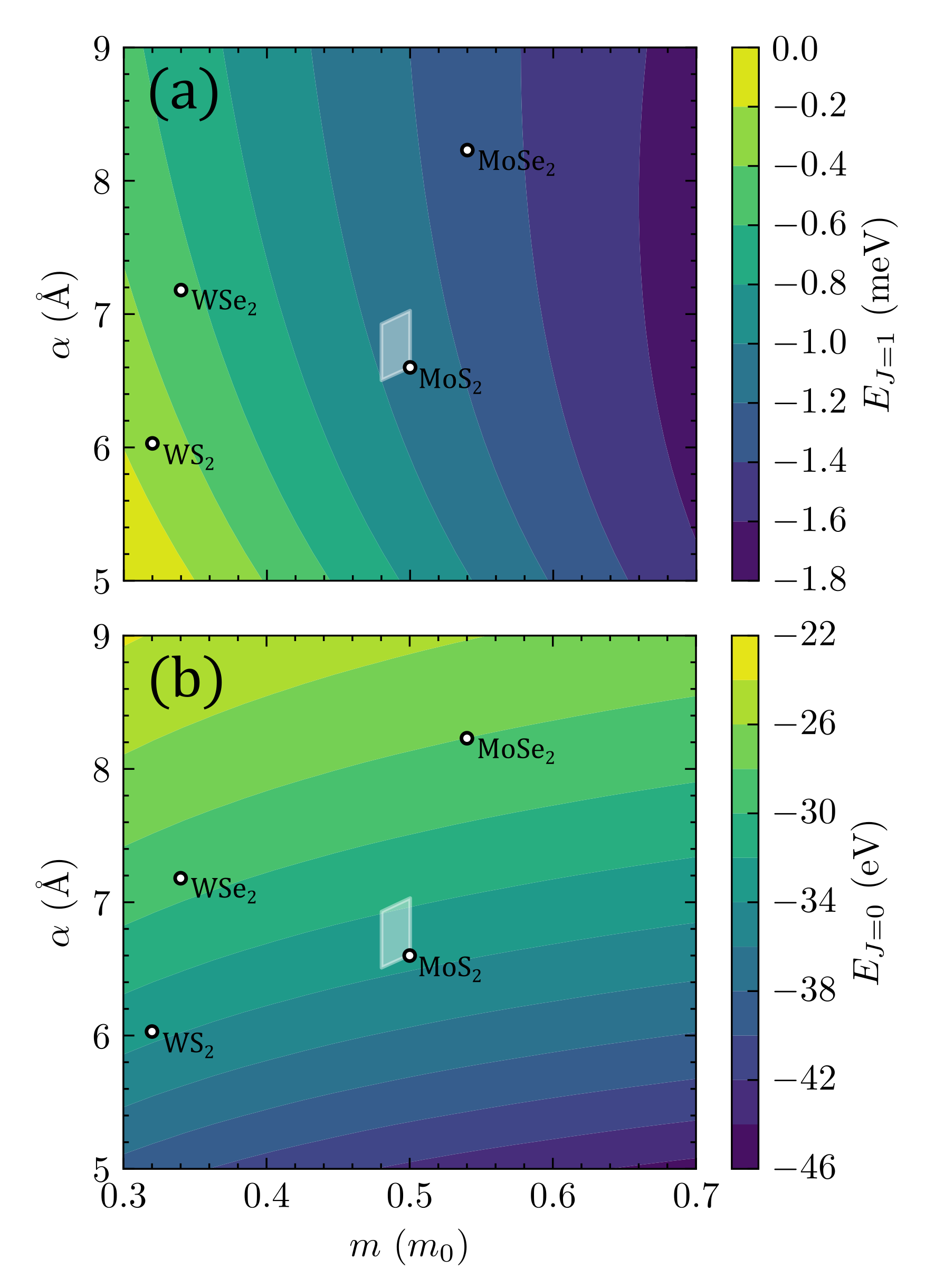}
  \caption{Trion binding energy landscape for  $J=1$ (top panel) and $J=0$ (bottom panel) states with variation in polarizability and charge carrier mass ($m_e=m_h$) in vacuum. The circles represents the different TMDs given in Table~\ref{tab:Table_Comp}. The box at the center of the figures estimate the MoS$_2$ trion binding energies  for small biaxial strain variations between 0\% and 2.0\% (see explanations in the text and Table~\ref{tab:model_params}). }
  \label{fig:color_map}
\end{figure}

\subsection{Strain effects}

We estimate the strain effect in the $J=0$ and $J=1$ trion  binding energies in suspended MoS$_2$, as a representative case for TMDCs. Strain experiments for MoS$_2$ monolayers (see Ref.~\cite{Strain_demonstration}) have shown that the optical absorption peaks shift while the exciton binding energies remain essentially unchanged, indicating a larger impact on the band structure. 
Another experiment report an increase in carrier mobility under strain~\cite{Strain_mobility}, consistent with theoretical predictions of a reduction in the electron and hole effective masses~\cite{Strain_mass}.

For modeling purposes, we assume electrons and holes share the same strain-dependent effective mass, restrict the strain to $\tau \leq 2\%$, which is below the onset of the direct–indirect bandgap transition~\cite{Strain_bandstructure}, and consider in this strain range a constant exciton binding energy~\cite{Strain_demonstration}. Under such assumptions, we estimate how trion binding energies responds to strain. 

For the unstrained and suspended MoS$_2$ monolayer we take the polarizability  and  effective mass $(m_e=m_h)$ from Ref.~\cite{Berkelbach2013} providing the value of the exciton binding energy of $E_{2B}=555.03$\,meV  (see Table~\ref{tab:Table_Comp}).  The polarizability is then tuned to keep fixed the exciton binding energy with strain~\cite{Strain_demonstration} as a function of the effective mass. We assume that $m_e$ decreases linearly up to 4\% as indicated  by ab-initio calculations computed for a biaxial 2\% strain as shown in Fig.~4a of Ref.~\cite{Strain_mass}. 
A decrease in effective mass enhances the kinetic energy contribution to the exciton energy. Therefore, maintaining a fixed binding energy requires a corresponding strengthening of the Coulomb attraction. In our effective model this dependence is encoded through the polarizability parameter $\alpha$, which controls the screening length. The resulting values of $\alpha$ as a function of $m$ are provided in Table~\ref{tab:model_params}.

Under the above assumptions, we compute the $X^-$ binding energies as a function of $\tau$ for both angular momentum states. We found a slight variation in the binding energies, namely an increase for the $J=0$ state and a decrease for $J=1$ trion, as shown in  Table~\ref{tab:model_params}. For the $J=0$ state the mass decrease that enhances the kinetic energy is almost balanced by the increase of the potential, as the polarizability diminishes, leading to a net enhancement of 0.2\% in the binding energy, for a $-4\%$ variation of the charge carriers masses and up to 2\% of strain. On the other hand, the weakly bound $J=1$ trion extends to large distances (see e.g. Fig.~\ref{fig:geomJ_panels}) where the charge carriers explore the tail of the  Rytova-Keldysh potential, which is weakly dependent on $\alpha$ (see Ref.~\cite{Cudazzo}), and thus the enhancement of the kinetic energy implies the lowering of the binding energy. Noteworthy to observe that in this case the binding is depleted roughly by 6\%, which is comparable to the enhancement of the kinetic energy due to the decrease of 4\% in the charge carrier masses. 

We have supposed the exact independence of the exciton binding energy for strains up to 2\% in order to estimate the effect on the trion states. However, as experimentally observed for suspended WSe$_2$ monolayer~\cite{Strain_tuning}, there is a decrease of about $25$ meV in the exciton binding for 2\% strain. We assume that a similar depletion occurs for the exciton in suspended MoS$_2$ monolayer and from that obtained the range of polarizability while keeping constant  $E_{2B}=530\,$meV, away from the unstrained value of 555\,meV (see Table~\ref{tab:Table_Comp}). Our calculations considered the interval of  charge carriers masses between 0.5$m_0$ and 0.48$m_0$, namely, up to 2\% strain. With these new set of parameters the binding energies were again computed for the $J=1$ and $J=0$ states and the results delimits the gray region seen in Fig.~\ref{fig:color_map} for our illustrative case.  For reference in the figure we also include the trion binding energies of MoSe$_2$, WS$_2$ and WSe$_2$ for the same parameters used in the SVM calculations~\cite{SVM_1} (see Table~\ref{tab:Table_Comp}).

We should comment about the different contour patterns observed in Fig.~\ref{fig:color_map} for the dependence of the $J=0$ and $J=1$ binding energies on $\alpha$ and charge carriers mass. The observed difference is due to the weak dependence of the large $J=1$ binding energy with $\alpha$ for fixed masses, while the binding of the compact $J=0$ state decreases by enhancing the polarizability, as the strength of the potential is depleted in this direction. In both cases, for a fixed polarizability, the binding energy increases by turning the charge carriers heavier, as the kinetic energy decreases. 

It is worthwhile to mention that the binding of the $X^+$ states is comparatively enhanced to the $X^-$ ones in realistic cases, where in general the hole is heavier than the electron, and therefore less kinetic energy. In Ref.~\cite{Strain_mass} the electron and hole masses from ab-initio calculations in suspended MoS$_2$ are provided in their Fig.~4a, for the unstrained case and for 2\% biaxial strain. Our results for the $X^+$ are presented in Table~\ref{tab:t+}, where we notice an expected enhancement mainly in the $J=1$ binding energy with respect to the equal mass case. The strain effect up to 2\% is almost irrelevant for the $J=0$ state binding energy, as already we have discussed for the $X^-$, while for $J=1$ the binding decreases by about 7\%, due to the increase of the kinetic energy with strain. Taking into account the decrease of the exciton binding by 25\,meV with the 2\% strain, there is a further depletion in the $J=0$ and $J=1$ binding, which is larger in the first case.

\begin{table}[t!]
  \caption{Estimative of the strain effect  in the binding energy of the $J=0$ and $J=1$ positive trion  states in suspended MoS$_2$ monolayer. We use the charge carrier masses parameters from the ab-initio calculations presented in Fig.~4a of Ref.~\cite{Strain_mass}.  The polarizability is  tuned to keep fixed the exciton binding energy with strain~\cite{Strain_Review} using as starting value the one quoted in Table~\ref{tab:Table_Comp}). In the last  line of the table the exciton binding is decreased by 25\,meV. }
  \label{tab:t+}
  \normalsize
  \resizebox{\columnwidth}{!}{
  \setlength{\tabcolsep}{6pt}
  \begin{tabular}{@{}c c c c c c c@{}}
    \toprule
    $\tau$ (\%) & $m_h$ ($m_0$) & $m_e$ ($m_0$) & $\alpha$ (\AA) & $E_{2B}$(meV)& $E_{J=0}$(meV)& $E_{J=1}$(meV) \\
    \midrule
    0.0\%  &  0.6  &  0.47 & 6.60 & 562.76 & 34.03 & 3.62  \\
    2.0\%  & 0.585 & 0.45 & 6.54 & 562.76  & 34.09 & 3.37  \\
    \midrule
    2.0\%  & 0.585 & 0.45 & 6.93  & 537.76 & 32.56 & 3.31  \\
    \bottomrule
  \end{tabular}
  }
\end{table}

\subsection{Dielectric Effects}

In this subsection we study the impact of mass imbalance and dielectric substrate in the binding energy of the $J=0$ and $J=1$ trion state. Furthermore, we follow the evolution of the $J=1$ state geometry as a function of these parameters. 

\begin{figure}[t!]
  \includegraphics[width=8cm]{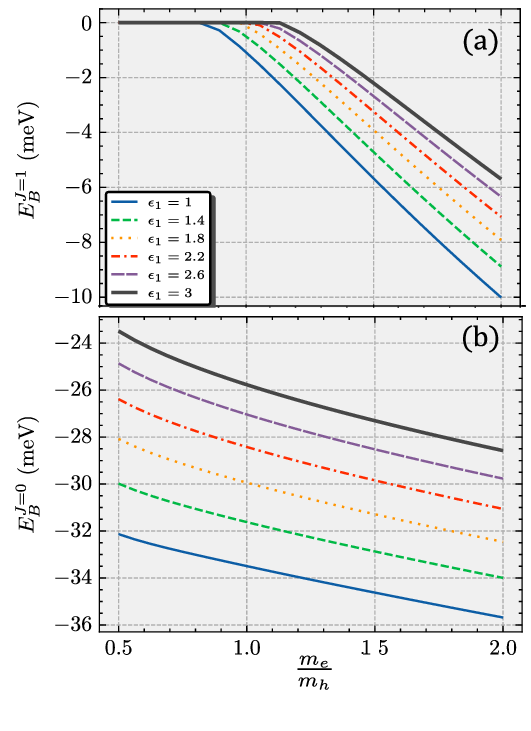}
  \caption{Dependence of the MoS$_2$ $J=0$ and $J=1$ state binding energy with respect to the ratio of the electron and hole effective masses for several dielectric constants. The hole effective mass is fixed to $m_h = 0.5\,m_0$, while the electron mass is varying. }
  \label{fig:fig_die_mass}
\end{figure}

The results for the  binding energy of the $J=1$ state as a function of the ratio $0.5<m_e/m_h<2$ for dielectric constants up to 3 are shown in panel (a) of Fig.~\ref{fig:fig_die_mass}, where $m_e/m_h<1$ and $m_e/m_h\geq 1$, corresponds to $X^{-}$ and $X^{+}$, respectively (cf. Table~\ref{tab:Table_Comp}). The $X^{+}$ is more bound than $X^{-}$ due to the decrease in the kinetic energy (see for reference Table~\ref{tab:t+}), clearly seen in the figure~\ref{fig:fig_die_mass}, by comparing the binding energy in the $m_e/m_h>1$ and $m_e/m_h\leq 1$ regions.

The increase of the dielectric constant weakens the RK potential in the asymptotic region, which decreases the $J=1$ binding up to a critical value of $\epsilon_1$ where the state is no longer bound. That happens both for $m_e/m_h\leq 1$ $(X^{-})$ and $>1$ $(X^{+})$, however the positive $J=1$ trion is more resilient to the increase of the dielectric constant. This is indeed expected, as the RK potential weakens by increasing the dielectric constant proportional to $1/\epsilon_1$ and to keep the state bound, it is necessary a reduction of the kinetic energy of the identical charge pair. Due to the very shallow nature of the $X^{-}$ bound-state, high dielectric environments yields a vanishing binding energy, while  $X^+$ still survives. In panel (b) of Fig.~\ref{fig:fig_die_mass} we present the dependence of the $J=0$ state binding energy as a function of the mass ratio $m_e/m_h$ for different dielectric substrates $\epsilon_1$. We observe that the binding energy tends to a plateau for $m_e/m_h \gg 1$ modulated by the dielectric substrate.

\begin{figure*}[t!]
  \includegraphics[width=1\linewidth]{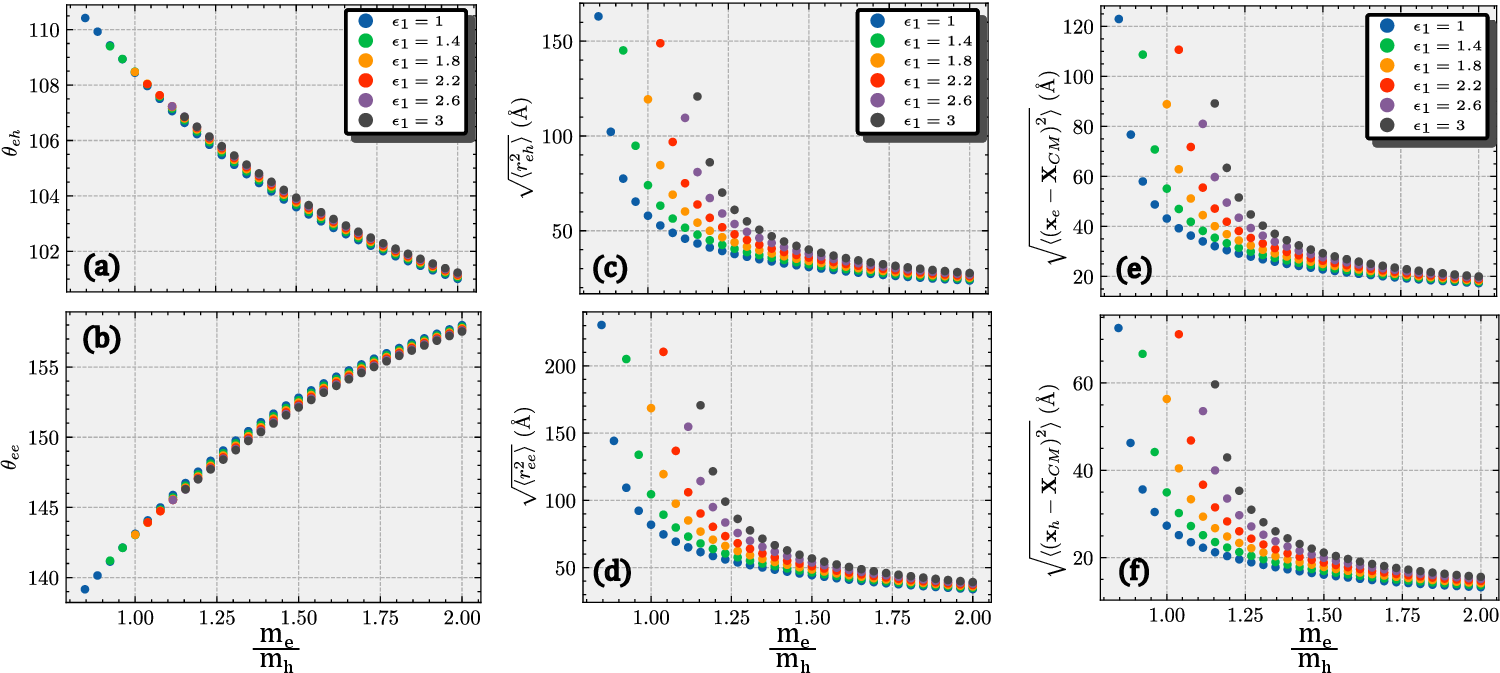}
  \caption{Dependence of the MoS$_2$  $X^-$  $J=1$  state geometric quantities with the dielectric constant, $\epsilon_1$, and electron effective mass. }
  \label{fig:fig_observables_dep}
\end{figure*}

The analysis of the $X^-$ and $X^+$ $J=1$ geometry is presented in Fig.~\ref{fig:fig_observables_dep}. The first striking property is seen in the dependence pf the angles with the mass imbalance, shown in panels (a) and (b), which are quite independent on the different dielectric environments. The increase of $\theta_{ee}$ with $m_e/m_h$ is consistent with the decrease of $\theta_{eh}$, signalizing the tendency of two heavy particles align with the center of mass of the system, which in the extreme case will correspond to $\theta_{ee}=180^\circ$ and $\theta_{eh}=90^\circ$. The trion sizes are studied in panels (c) and (d), where it is shown the root mean square relative distances between the charge carriers, i.e., $\sqrt{\langle r^2_{eh}}\rangle$ and $\sqrt{\langle r^2_{ee}}\rangle$, respectively. Comparing both cases, the relative distance between the equally charged carriers is larger than the one for opposite charges, due to the repulsive RK potential in the first case. The increase of the dielectric constant depletes the $J=1$ binding and consequently the sizes are enhanced, and of course, dramatically when the state is close to the critical situation of being no longer bound. The root mean square relative separation distances between the charge carriers tend to a plateau towards large mass imbalance $m_e/m_h\gg 1$, modulated by the dielectric constant. The same behavior is found for the root mean square separation distance of the electron and hole to the trion center of mass, presented in panels (e) and (f), respectively. The plateau indicates the limit of a system composed by two heavy identical particles and a light one. In this limit, the electrons are farther from the center of mass, while the hole is closer, which is a consequence of the repulsive interaction. All these calculations allow to conclude that the separation angle between the identical particles increases from $X^-$ to $X^+$, while the opposite behavior is valid for the angle between the opposite charge carriers. In addition, the size decreases from $X^-$ to the $X^+$, and correspondingly the binding increases. Such trends are independent on the dielectric constant. However, by tuning the dielectric constant one can find  a critical value where above it only the  $X^+$ survives for $J=1$, and the strain variation up to 2\% keeps this observation unchanged.

\section{Conclusion and perspectives}\label{sec:conclusion}

We have investigated the properties of trions in transition metal dichalcogenides (TMDC) monolayers using the Gaussian Expansion Method (GEM), suitably adapted to two-dimensional systems. Excitons and trions in monolayers of TMDCs with chemical composition MX\(_2\) in the 2H phase were studied systematically. We computed the binding energies of both excitons and trions, and identified, in addition to the known \( J=0 \) trion, the existence of a bound excited trion state with orbital angular momentum \( J=1 \).  We analyzed the sensitivity of the $J=1$ state to realistic environmental effects, such as strain and dielectric screening, illustrated for the MoS$_2$ case~\cite{chaves2020bandgap,raja2017coulomb,Strain_tuning}. We showed that small strain, up to 2\% mainly affects the $J=1$ state through effective-mass renormalization, leading to a slight reduction of its binding energy, while the compact $J=0$ trion remains essentially unchanged \cite{Strain_demonstration,Strain_mass}. In contrast, dielectric screening strongly weakens the long-range Rytova–Keldysh interaction, driving the $J=1$ $X^{-}$ trion to the continuum for large enough dielectric substrate constant, while the $X^{+}$ counterpart is more resilient to such environment modifications owing to the fact that the hole is typically heavier than the electron. 

Our results for the \( J=0 \) trion are compared against previous calculations using the Stochastic Variational Method (SVM), Quantum Monte Carlo (QMC) and many others, showing excellent agreement. Furthermore, we examined the internal structure and geometry of the trion states through their probability density distributions and root-mean-square separations. We find that the spatial configuration of the \( J=1 \) trion is qualitatively similar to that of the \( J=0 \) state, though it is stretched by a factor of approximately 2.5 to 3 in size, leading to sizes from to 5 to 10 nm. This geometric difference, along with its lower binding energy, suggests that the \( J=1 \) trion may have a significantly longer lifetime and could potentially be observed in cryogenic optical experiments, given its binding energy of order 1\,meV below the exciton for the $X^-$, and about 3\,meV for the $X^+$. Moreover examining its symmetry to electron (hole) exchange in $X^-$ ($X^+$), we discussed if the state is bright or dark, given a symmetric spin-valley configuration. We emphasized that the singlet-singlet configuration of the $J=1$ state may turn it a possible candidate bright state, while the triplet-triplet configuration is blocked to recombine due to spin/valley mismatches, yielding a dark state. Although, one has to remind that our effective Hamiltonian is limited with respect to many-body effects. On the other side, recent experimental advances on the coupling between TMDCs with plasmonic nanocavities show a dramatic enhancement of the trion emission peak in WS$_2$-Ag nanocavity~\cite{nanocav_1} and in WSe$_2$-Au nanocavity~\cite{nanocav_2}. Beyond that, more exotic states from a new kind of cooperative coupling between plasmon and different excitonic complexes presenting a large Bohr radius of up to 5 nm, namely the plasmon-$XX^{-}$, was recently observed in WS$_2$-silver nanocavities~\cite{nanocav_3}. The possibility to enhance the emission peaks of different excitonic complexes in TMDCs coupled to nanocavities is an exciting avenue to probe the few-body description of such two-dimensional states. 

The GEM, long established in nuclear and atomic physics for the study of strongly interacting few-body systems~\cite{hiyama_example_5,Tetramer_GEM,Hiyama2012_2}, proves to be a highly efficient and accurate method for calculating  weakly bound few-body bound states in two-dimensional materials. GEM can also be combined with the Complex Scaling Method to calculate resonances~\cite{Resonance_1,Resonance_2} or, combining with the Kohn-Type variational principle to compute the S-Matrix, describe reactions~\cite{Reaction_1} and scattering~\cite{Hiyama2006}. Furthermore, given the large variety of applications of GEM in three-dimensional systems in different contexts, the extensions to systems of four and five-body excitonic complexes in lower dimensions, should not pose severe complications due to simplifications concerning angular momentum algebra and the dominance of central interactions. In the near future, we envisage the exciting exploration with GEM of excitonic complexes in multilayered systems~\cite{geim2013van,li2016heterostructures,exp_t2trion,exp1_t1trion,exp2_t1trion} and anisotropic semiconductors~\cite{Phosp_1,Phosp_2,Phosp_3}, which offer a multitude of possibilities and combinations within the context of modern applications in optoelectronics~\cite{geim2013van,Phosp_5,Phosp_6}.

\section{Acknowledgments} 
\thispagestyle{plain}
E.H. and L.T. are supported by ERATO-JPMJER2304. E.H. is also supported by Kakenhi-23K03378 and Grant-in-aid for Transformative Research Areas (A)-25H01268. T.F. is supported in part by the National Council for Scientific and Technological Development (CNPq) (Grant No. 306834/2022-7) and S\~ao Paulo Research Foundation (FAPESP) (Grant No. 2023/13749-1 and 2024/17816-8). A.~J.~C. acknowledge CNPq grants No. 423423/2021-5, 408144/2022-0, and 315408/2021-9.  We thank the financial support from  Improvement of Higher Education Personnel CAPES (Finance Code 001). This work is a part of the project Instituto Nacional de  Ci\^{e}ncia e Tecnologia - F\'{\i}sica Nuclear e Aplica\c{c}\~{o}es Proc. No. 464898/2014-5.
\appendix

\section{ Matrix Elements for GEM} \label{app:matelem}

\subsection{Two-Body problem}\label{app:matelemB}
The secular equation for the exciton problem reads:
\begin{equation}
   \sum_{n^{\prime}} \left(H^{m}_{n,n^{\prime}} - E\,N^{m}_{n,n^{\prime}}\right)\mathcal{A}^{m}_{n,n^{\prime}} = 0\,,
\end{equation}
where
\begin{align}
   H^{m}_{n,n^{\prime}} &= \bra{\phi_{n,m}}H\ket{\phi_{n^{\prime},m}}, \\ 
   N^{m}_{n,n^{\prime}} &= \bra{\phi_{n,m}}1\ket{\phi_{n^{\prime}m}}.
\end{align}

The Hamiltonian matrix elements, $H^{m}_{n,n^{\prime}} = T^{m}_{n,n^{\prime}} + V^{m}_{n,n^{\prime}}$ are given by 
\begin{align}
 T^{m}_{n,n^{\prime}} & = \mathcal{C}^m_{n,n'}\frac{\hbar^2\pi\nu_n \nu_{n^{\prime}}\Gamma\left(2+\lvert m\rvert\right)}{\mu_X\left(\nu_n + \nu_{n^{\prime}}\right)^{\lvert m\rvert+2}}, \\
V^{m}_{n,n^{\prime}} &=\mathcal{C}^m_{n,n'}\int d\mathbf{r} r^{2\lvert m\rvert+1} e^{-\left(\nu_{n}+\nu_{n^{\prime}}\right) r^2}V(r)\,, 
\end{align}
where 
\begin{equation}
\mathcal{C}^m_{n,n'}=N_{n,\lvert m\rvert}N_{n^{\prime},\lvert m\rvert}\,,
\end{equation}

The basis functions overlap is given by:
\begin{align}
N^{m}_{n,n^{\prime}} & = N_{n,\lvert m\rvert}N_{n^{\prime},\lvert m\rvert}\frac{\Gamma\left(\lvert m\rvert + 1\right)}{2\left(\nu_n + \nu_{n^{\prime}}\right)^{\lvert m\rvert + 1}}.
\end{align}

\subsection{Three-Body Matrix Elements}\label{app:matelemC}

The matrix elements for each operator appearing in the three-body relative Hamiltonian~\eqref{eq:Red_Ham} is provided in what follows. The analytical details and further discussion regarding the infinitesimally shifted Gaussian-lobe (ISGL) function are presented in the supplemental material. 
In the ISGL representation of the basis functions the overlap between  them are:
\begin{multline}
    \mathcal{N}^{(i j)}_{\alpha_i\alpha_j} 
    = \frac{\pi^2 \bar{N}}{\zeta^{ij}_j\eta^{\prime\,ij}_j
        \nu^{\lvert \ell_i\rvert}_{n_i}
        \nu^{\lvert \ell_j\rvert}_{n_j}
        \lambda^{\lvert L_i\rvert}_{N_i}
        \lambda^{\lvert L_j\rvert}_{N_j}} \\ \hspace{-.2cm}\times 
    \sum_{n=1}^{n_{\text{max}}}
    S^{J,n}_{\ell_{i},\ell_{j},L_{i},L_{j}}
    \mathcal{F}^{a^{(n)}_{12}}_{12}
    \mathcal{F}^{a^{(n)}_{13}}_{13}
    \mathcal{F}^{a^{(n)}_{14}}_{14}
    \mathcal{F}^{a^{(n)}_{23}}_{23}
    \mathcal{F}^{a^{(n)}_{24}}_{24}
    \mathcal{F}^{a^{(n)}_{34}}_{34},
\end{multline}
 where  $a^{(n)}_{ij}$ are exponents throughout this appendix,   $\bar{N} = N_{n_i,\lvert \ell_i\rvert}N_{N_i,\lvert L_i\rvert}N_{n_j,\lvert \ell_j\rvert}N_{N_j,\lvert L_j\rvert}$ and $\alpha_i\equiv \{n_i,N_i,\ell_i,L_i\}$. The quantities $a^{(n)}_{ij}$ and $S^{J,n}_{\ell_{i},\ell_{j},L_{i},L_{j}}$ are  defined in the supplemental material.  $ \mathcal{F}_{ij} $ is given by:
\begin{equation}
    \mathcal{F}_{ij} = \frac{2p_ip_j}{\zeta^{ij}_{j}} + \frac{2q_iq_j}{\eta^{\prime \, ij}_{j}},
\end{equation}
with the definitions:
\begin{align}
p_1& = \nu_{n_i}\, \sigma_{ic}\,, \nonumber\\
p_2 &= \lambda_{N_i}\, \beta_{ic}\,,\nonumber\\
p_3& = \nu_{n_j\,} \sigma_{jc}\,,\nonumber\\
p_4 &= \lambda_{N_j}\, \beta_{jc}\,,
\\ \nonumber \\
q_1  &= \nu_{n_i} \gamma_{ic} - f^{ij}_{c} \,p_1\,, \nonumber\\
q_2 &= \lambda_{N_i} \alpha_{ic} - f^{ij}_{c} \,p_2\,,\nonumber \\
q_3 &= \nu_{n_j} \gamma_{jc} - f^{ij}_{c}\, p_3, \nonumber\\
q_4 &= \lambda_{N_j} \alpha_{jc} - f^{ij}_{c} p_4\,,
\end{align}
where \( f^{ij}_{c} = \xi^{ij}_c / \zeta^{ij}_c \) and the factors for the transformation between rescaled channel Jacobi relative coordinates,
\begin{eqnarray}
&&\xi^{ab}_c = \nu_a \,\gamma_{ac} \,\sigma_{ac} + \lambda_a\, \alpha_{ac} \,\beta_{ac} + \nu_b \,\gamma_{bc}\, \sigma_{bc} + \lambda_b\, \alpha_{bc} \,\beta_{bc}, \nonumber\\
&&\eta^{ab}_c = \nu_a \,\gamma_{ac}^2 + \lambda_a \,\alpha_{ac}^2 + \nu_b \,\gamma_{bc}^2 + \lambda_b \,\alpha_{bc}^2, \nonumber \\
&&\zeta^{ab}_c = \nu_a \,\sigma_{ac}^2 + \lambda_a \,\beta_{ac}^2 + \nu_b\, \sigma_{bc}^2 + \lambda_b \,\beta_{bc}^2,  
\end{eqnarray}
with $\eta^{\prime \, ab}_{c} = \eta^{ab}_c - \xi^{ab}_c/\zeta^{ab}_c$ and $c = b$ for the case of the overlap matrix elements. The factors $\alpha_{ac},\beta_{ac},\gamma_{ac}$ and $\sigma_{ac}$ are obtained by solving the following system of equations,
\begin{align}
    \mathbf{r}_{a} & = \gamma_{ac}\mathbf{r}_c+ \sigma_{ac}\mathbf{R}_c, \\ 
    \mathbf{R}_{a} & = \alpha_{ac}\mathbf{r}_c + \beta_{ac}\mathbf{R}_c.
\end{align}
The matrix elements of the potential reads:
\begin{multline}
\mathcal{V}^{(ij,c)}_{\alpha_i\alpha_j} =\frac{2\pi^2 \bar{N}}{\zeta^{ij}_{c}\nu^{\lvert \ell_i\rvert}_{n_i}\nu^{\lvert \ell_j\rvert}_{n_j}\lambda^{\lvert L_i\rvert}_{N_i}\lambda^{\lvert L_j\rvert}_{N_j}} 
\\  \times\sum_{n=1}^{n_{max}}S^{J,n}_{\ell_{i},\ell_{j},L_{i},L_{j}}\sum_{k=0}^{M_{sum}}w_{k}\,\Lambda^{(n)}_{V,k}\,,    
\end{multline}
where,
\begin{equation}
\Lambda^{(n)}_{V,k}=\mathcal{G}^{a^{(n)}_{12}}_{12,k}\,\mathcal{G}^{a^{(n)}_{13}}_{13,k}\mathcal{G}^{a^{(n)}_{14}}_{14,k}\,\mathcal{G}^{a^{(n)}_{23}}_{23,k}\,\mathcal{G}^{a^{(n)}_{24}}_{24,k}\,\mathcal{G}^{a^{(n)}_{34}}_{34,k}\,,
\end{equation}
and
\begin{equation}
    \mathcal{G}_{ij,n} = \frac{2p_i p_j}{\zeta^{ij}_{c}} + \frac{2\mu_n q_i q_j}{\eta^{\prime \,ij}_{c}}\,.
\end{equation}
The matrix elements of the kinetic energy operator are:
\begin{multline} \mathcal{K}^{(ij)}_{\alpha_i\alpha_j} =\frac{\pi^2 \bar{N}}{\eta^{\prime \, ij}_{j}\zeta^{ij}_{j}\nu^{\lvert \ell_i\rvert}_{n_i}\nu^{\lvert \ell_j\rvert}_{n_j}\lambda^{\lvert L_i\rvert}_{N_i}\lambda^{\lvert L_j\rvert}_{N_j}} 
\\ \times \sum_{n=1}^{n_{max}}S^{J,n}_{\ell_{i},\ell_{j},L_{i},L_{j}}\sum_{k=0}^{M_{sum}}w_{k}\,\Lambda^{(n)}_{K,k}\,,
\end{multline}
where,
\begin{equation} \Lambda^{(n)}_{K,k}=t^{a^{(n)}_{12}}_{12,k}\,t^{a^{(n)}_{13}}_{13,k}\,t^{a^{(n)}_{14}}_{14,k}\,t^{a^{(n)}_{23}}_{23,k}\,t^{a^{(n)}_{24}}_{24,k}t^{a^{(n)}_{34}}_{34,k}\,,
\end{equation}
with 
\begin{equation}
   t_{ij,n} = \frac{2p_ip_j}{\zeta^{ij}_{j}}+\frac{2q_iq_j}{\eta^{\prime \, ij}_{j}}+\frac{\mu_n K_{ij}}{K_0}
\end{equation}
and
\begin{equation}
K_{0} =
4\,\tau_{r_j}\,\nu_{j}\!\left(\frac{\nu_{j}}{\eta^{\prime \,ij}_j}-1\right)
  + 4\,\tau_{R_j}\,\lambda_{j}\!\left(\frac{\lambda_{j}}{\zeta^{\prime \,ij}_j}-1\right),     
\end{equation}
\begin{eqnarray}
&&K_{mn} =\frac{4\,\tau_{r_n}\,\nu_{j}^{2}}{\eta^{\prime \,mn}_n}
  \Bigl(
      q_{m}q_{n}
      - q_{1}\delta_{m,1}\delta_{n,3}
      - q_{2}\delta_{m,2}\delta_{n,3} \nonumber\\
      &&~~~ - q_{4}\delta_{m,3}\delta_{n,4}
  \Bigr)
+
  \frac{4\,\tau_{R_n}\,\lambda_{j}^{2}}{\zeta^{\prime \,mn}_n}\!
  \Bigl(
      q'_{m}q'_{n}
      - q'_{1}\delta_{m,1}\delta_{n,3}\nonumber\\
      && \hspace{2.5cm} - q'_{2}\delta_{m,2}\delta_{n,3}
      - q'_{4}\delta_{m,3}\delta_{n,4}
  \Bigr),
\end{eqnarray}
where for the kinetic energy matrix elements,
\begin{align*}
p_1 &= \nu_a \sigma_{ab}, & 
p_2 &= \lambda_a \beta_{ab}, \\\
p_3 &= 0, & 
p_4 &= \lambda_b, \\
p_1' &= \nu_a \gamma_{ab}, & 
p_2' &= \lambda_a \alpha_{ab}, \\
p_3' &= \nu_b, & 
p_4' &= 0\,.
\end{align*}
The factors $\tau_{r_n}$ and $\tau_{R_n}$ are defined as,
\begin{equation}
    \tau_{r_n} = \frac{\hbar^2}{2 \mu_{r_n}} \quad\text{and}\quad
    \tau_{R_n} = \frac{\hbar^2}{2 \mu_{R_n}}\,.
\end{equation}

\bibliographystyle{apsrev4-2}
\thispagestyle{plain}
\bibliography{references}
\thispagestyle{plain}

%% file: supp_body.tex


\title{Supplemental Material: Gaussian Expansion Method for the study of few-body states in two-dimensional materials. }	

\author{Luiz G. M. Ten\'orio\orcidlink{}}
\email{luiztenorio@fisica.ufmt.br}
\affiliation{Department of Physics, Graduate School of Science, Tohoku University, Sendai 980-8578, Japan}
\affiliation{Instituto Tecnológico de Aeronáutica, 12.228-900, São José dos Campos, SP, Brazil} 

\author{André J. Chaves\orcidlink{}}
\affiliation{Instituto Tecnológico de Aeronáutica, 12.228-900, São José dos Campos, SP, Brazil} 
\author{Emiko Hiyama\orcidlink{0000-0002-6352-5766}}
\email{hiyama@riken.jp}
\affiliation{Department of Physics, Graduate School of Science, Tohoku University, Sendai 980-8578, Japan}
\affiliation{RIKEN Nishina Center for Accelerator-Based Science, Wako, Saitama 351-0198, Japan}

\author{ Tobias Frederico\orcidlink{0000-0002-5497-5490}
} 
\email{tobias@ita.br}

\affiliation{Instituto Tecnológico de Aeronáutica, 12.228-900, São José dos Campos, SP, Brazil} 

\date{\today}
\maketitle
\section{Gaussian Expansion Method}
\subsection{Infinitesimally Shifted Gaussian-Lobe Function}

In this section, we derive the infinitesimally shifted Gaussian-lobe representation of the basis function. The representation applies to both three- and two-dimensional calculations, so we will present the more general three-dimensional case and then reduce it to the two-dimensional case. The construction follows ref.~\cite{hiyama_GEM} and will be presented here for completeness.

For the purpose of the construction of the representation, consider the following function parametrized by three integers, $n,\ell$ and $m$, and a position vector $\mathbf{r}$,
\begin{equation}\label{eq:SM1}
    \phi_{n,\ell,m}(\mathbf{r}) = N_{n,\ell} r^{\ell} e^{-\nu_n r^2} Y_{\ell,m}(\hat{\mathbf{r}}),
\end{equation}
where $N_{n,\ell}$ is a normalization constant, $\nu_n$ is a positive real number and $Y_{\ell,m}(\hat{\mathbf{r}})$ is a spherical harmonic function. Here, we will restrict ourselves to $\ell >0$ and $m>0$. If one considers systems of three- or more-bodies, a matrix element would involve a large number of coupled spherical harmonics, which would introduce complicated Racah algebra and yield very laborious calculations. Thus, to simplify the problem, we introduce the following representation, called the Infinitesimally Shifted Gaussian-Lobe (ISGL) representation, 
\begin{equation}\label{eq:SM2}
    N_{n,\ell} r^{\ell} e^{-\nu_n r^2} Y_{\ell,m}(\hat{\mathbf{r}}) = N_{n,\ell}\lim_{\epsilon \to 0}\frac{1}{(\nu_n \epsilon)^{\ell}}\sum_{k=1}^{k_{max}}C_{\ell m,k}e^{-\nu\left(\mathbf{r}-\epsilon \mathbf{D}_{\ell m,k}\right)^2},
\end{equation}
where the parameters $C_{\ell m,k}$ and $\mathbf{D}_{\ell m,k}$ are called Shift Parameters (SP). First, we will derive the representation, SPs, and then we will discuss the intuition behind the ISGL representation.  Starting with the Cartesian representation of $r^{\ell}Y_{\ell m}$:
\begin{align}\label{eq:SM3}
    r^{\ell}Y_{\ell m} & = \left(\frac{(2\ell+1)(\ell-m)!}{4\pi (\ell+m)!}\right)^{\frac{1}{2}}\frac{(\ell+m)!}{2^{m}} r^{\ell} \left(\sin \theta e^{i\phi}\right)m\sum_{j=0}^{\left[\frac{\ell-m}{2}\right]}\frac{\left(-1\right)^{j}\cos^{\ell-m-2j}\theta \sin^{2j}\theta}{4^{j}j!(m+j)!(l-m-2j)!} \\
    &=\sum_{j=0}^{\frac{\ell-m}{2}}A_{\ell m,j}z^{\ell-m-2j}(x+iy)^{m+j}(x-iy)^{j}
\end{align}
where,
\begin{equation}\label{eq:SM4}
    A_{\ell m,j} =  \left(\frac{(2\ell+1)(\ell-m)!}{4\pi (\ell+m)!}\right)^{\frac{1}{2}}\frac{(\ell+m)!}{2^{m}}\frac{\left(-1\right)^{j}}{4^{j}j!(m+j)!(\ell-m-2j)!}.
\end{equation}

We introduce the following vectors:
\begin{align}\label{eq:SM5}
    \mathbf{a}_{z} &= (0,0,1), \\
    \mathbf{a}_{xy} &= (1,i,0), \\
    \mathbf{a}^{*}_{xy} &= (1,-i,0),
\end{align}
which allows us to write Eq.~\eqref{eq:SM3} as,
\begin{equation}\label{eq:SM6}
   r^{\ell}Y_{\ell m} = \sum_{j=0}^{\frac{\ell-m}{2}}A_{\ell m,j}(\mathbf{a}_{z}\cdot \mathbf{r})^{\ell-m-2j}(\mathbf{a}_{xy}\cdot \mathbf{r})^{m+j}(\mathbf{a}^{*}_{xy}\cdot \mathbf{r})^{j}.
\end{equation}
Next, we apply the following identity,
\begin{align}\label{eq:SM7A}
    \mathbf{a}\cdot\mathbf{r} &= \lim_{\epsilon \to 0} \frac{1}{4\epsilon \nu_n}\left[e^{2\epsilon\nu_n \mathbf{a}\cdot \mathbf{r}}-e^{-2\epsilon\nu_n \mathbf{a}\cdot \mathbf{r}}\right], \\ 
    &=\lim_{\epsilon \to 0} \frac{1}{4\epsilon \nu_n}\left[4\epsilon \nu_n \mathbf{a}\cdot \mathbf{r} + \mathcal{O}(\epsilon^{2})\right],
\end{align}
to Eq.~\eqref{eq:SM6}, yielding,
\begin{align}
    r^{\ell}Y_{\ell m}&=\sum_{j=0}^{\frac{\ell-m}{2}}A_{\ell m,j}\lim_{\epsilon\to 0}\left(\frac{1}{4\nu_n \epsilon}\right)^{\ell}\left[e^{2\epsilon\nu_n \mathbf{a}_{z}\cdot \mathbf{r}}-e^{-2\epsilon\nu_n \mathbf{a}_{z}\cdot \mathbf{r}}\right]^{\ell-m-2j}\left[e^{2\epsilon\nu_n \mathbf{a}_{xz}\cdot \mathbf{r}}-e^{-2\epsilon\nu_n \mathbf{a}_{xz}\cdot \mathbf{r}}\right]^{m+j}\times \nonumber \\ 
    &\times\left[e^{2\epsilon\nu_n \mathbf{a}^{*}_{xz}\cdot \mathbf{r}}-e^{-2\epsilon\nu_n\mathbf{a}^{*}_{xz}\cdot \mathbf{r}}\right]^{j}.
\end{align}
This expression can naturally be expanded using the binomial expansion, 
\begin{align}\label{eq:SM7B}
   & r^{\ell}Y_{\ell m}=\sum_{j=0}^{\frac{\ell-m}{2}}A_{\ell m,j}\lim_{\epsilon\to 0}\left(\frac{1}{4\nu_n \epsilon}\right)^{\ell}\, \sum_{s=0}^{\ell - m -2j}\binom{\ell - m -2j}{s}(-1)^{\ell-m-2j-s}e^{2(2s-\ell +m +2j)\epsilon\nu_n\mathbf{a}_z\cdot \mathbf{r}} \nonumber \\
    &\times \sum_{t=0}^{j+m}\binom{j+m}{t}(-1)^{j+m-t}e^{2(2t-j-m)\epsilon\nu_n\mathbf{a}_{xy}\cdot \mathbf{r}} \sum_{u=0}^{j}\binom{j}{u}(-1)^{j-u}e^{2(2u-j)\epsilon\nu_n\mathbf{a}^{*}_{xy}\cdot \mathbf{r}} \nonumber \\
    &= \sum_{j=0}^{\frac{\ell-m}{2}}A_{\ell m,j}\sum_{s=0}^{\ell - m -2j}\lim_{\epsilon\to 0}\left(\frac{1}{4\nu_n \epsilon}\right)^{\ell}\sum_{t=0}^{j+m}  \sum_{u=0}^{j}\binom{\ell - m -2j}{s}\binom{j+m}{t}\binom{j}{u}(-1)^{\ell-s-t-u}e^{2\nu_n\epsilon\mathbf{D}_{\ell m,k}\cdot \mathbf{r}},
\end{align}
where we conveniently define,
\begin{eqnarray}\label{eq:SM8}
    \mathbf{D}_{\ell m,k} =& D_x \hat{\mathbf{x}}+D_y \hat{\mathbf{y}}+D_z \hat{\mathbf{z}}, \\ 
    D_x =&2t+2u -2j-m ,\\ 
    D_y =&2t-2u-m ,\\ 
    D_z =&2s-\ell+m+2j.
\end{eqnarray}
Then, we can now write,
\begin{align}\label{eq:SM9}
    N_{n,\ell}\,r^{\ell} e^{-\nu r^2}Y_{\ell m} &= N_{n,\ell} \sum_{j=0}^{\frac{\ell-m}{2}}A_{\ell m,j}\lim_{\epsilon\to 0}\left(\frac{1}{4\nu_n \epsilon}\right)^{\ell}\sum_{s=0}^{\ell - m -2j}\sum_{t=0}^{j+m}  \sum_{u=0}^{j}\binom{\ell - m -2j}{s}\binom{j+m}{t}\binom{j}{u},\nonumber \\
    &\times (-1)^{\ell-s-t-u}e^{-\nu r^2+2\nu_n\epsilon\mathbf{D}_{\ell m,k}\cdot \mathbf{r}},
\end{align}
finally, the multi-index summation can be cast in a one-dimensional sum over $k$,
\begin{equation}\label{eq:SM10}
    N_{n,\ell}\,r^{\ell} e^{-\nu r^2}Y_{\ell m} = N_{\ell,m}\lim_{\epsilon \to 0}\left(\frac{1}{4\nu_n \epsilon}\right)^{\ell}\,\sum_{k=1}^{k_{max}}C_{\ell m,k}e^{-\nu_n \left(\mathbf{r}-\epsilon \mathbf{D}_{\ell m,k}\right)^2}.
\end{equation}
Equation~\eqref{eq:SM10} is the ISGL representation of the Gaussian Expansion Method's basis functions. In the matrix elements calculations, some very interesting properties emerge, such as the leading order in $\epsilon$ of the summation over $k$ is $\epsilon^{\ell}$ and any higher order terms is canceled as $\epsilon\to 0$. This allows for a analytical calculation of matrix elements with no round-off error induced by the limiting process as all $\epsilon$ factors are eliminated analytically. If we impose $\ell = \lvert m \rvert$ and $\theta = \pi/2$, the shift parameters in the $z$ direction are all identically zero and all harmonics are restricted to the plane, which yields the 2D representation of the ISGL basis functions. For negative $m$ quantum number, we use the fact $Y_{\ell,-m}(\hat{\mathbf{r}})  =(-1)^{m} Y_{\ell,m}(\hat{\mathbf{r}})$, therefore,
\begin{equation}
    C_{\ell -m,k} = (-1)^{m}C_{\ell m,k},
\end{equation}
\begin{equation}
    \mathbf{D}_{\ell -m,k} = \mathbf{D}^{*}_{\ell m,k}.
\end{equation}
\subsection{Three-Body Calculations}

In this section we present in detail how to calculate each matrix element for the three-body problem. 
The basis function in some channel $c$ is constructed using the infinitesimally shifted gaussian-lobe representation~\eqref{eq:SM10},
\begin{eqnarray}\label{eq:SM11}
    \Phi^{c}_{n_c,\ell_c,N_c,L_c} = N_{n_c,\lvert \ell_c\rvert}N_{N_c,\lvert L_c\rvert}\lim_{\epsilon_1,\epsilon_2 \to 0}\frac{1}{(4\nu_n \epsilon_1)^{\lvert \ell_c \rvert}}\frac{1}{(4\lambda_N \epsilon_2)^{\lvert L_c \rvert}}\sum_{k = 1}^{k_{max}}\sum_{K = 1}^{K_{max}}C_{\ell_c,k}C_{L_c,K} \nonumber\\ 
    \times e^{-\nu_n \left(\mathbf{r}_c-\epsilon_{1} \mathbf{D}_{\ell m,k}\right)^2}e^{-\lambda_n \left(\mathbf{R}_c-\epsilon_{2} \mathbf{D}_{L,K}\right)^2}.
\end{eqnarray}

The matrix elements for a given operator $\hat{F}$ consists of calculations of the following form,
\begin{equation}\label{eq:SM12}
\bra{\Phi_{c=1}+\Phi_{c=2}+\Phi_{c=3}}\hat{F}\ket{\Phi_{c=1}+\Phi_{c=2}+\Phi_{c=3}} = \sum_{i,j =1}^{3} \bra{\Phi_{i}}\hat{F}\ket{\Phi_j}.
\end{equation}

We begin with interaction potential matrix elements in some channel $c$:
\begin{equation}\label{eq:SM13}
\bra{\Phi_{i}}V(\hat{r}_c)\ket{\Phi_j} = \mathcal{V}^{(ij,c)}_{n_i,N_i,\ell_i,L_i;n_j,N_j,\ell_j,L_j},
\end{equation}
where:
\begin{equation}\label{eq:SM14}
\mathcal{V}^{(ij,c)}_{n_i,N_i,\ell_i,L_i;n_j,N_j,\ell_j,L_j} = \bar{N} \lim_{\epsilon^{1},\epsilon^{2},\epsilon^{3},\epsilon^{4} \to 0}\sum_{k_1,k_2,k_3,k_4}\frac{C_{\ell_{i},k_{1}}C_{L_{i},k_{2}}C_{\ell_{j},k_{3}}C_{L_{j},k_{4}}}{(\nu_{n_{i}}\epsilon_1)^{\lvert \ell_{i}\rvert }(\lambda_{N_{i}}\epsilon_{2})^{\lvert L_{i}\rvert}(\nu_{n_{j}}\epsilon_3)^{\lvert \ell_{j}\rvert}(\lambda_{N_{j}}\epsilon_{4})^{\lvert L_{j}\rvert}} I^{ij}_{c,V},
\end{equation}
\begin{equation}\label{eq:SM15}
    I^{ij}_{c,V}=\int d\mathbf{r}_{c}d\mathbf{R}_{c}e^{-\nu_n(\mathbf{r}_i-\epsilon_{1} \mathbf{D}_{\ell_{i},k_{1}})^2}e^{-\lambda_{N_{i}}(\mathbf{R}_i-\epsilon_{2} \mathbf{D}_{L_{i},k_{2}})^2}e^{-\nu_{n_{j}}(\mathbf{r}_j-\epsilon_{3} \mathbf{D}_{\ell_{j},k_{3}})^2}e^{-\lambda_{N_{j}}(\mathbf{R}_j-\epsilon \mathbf{D}_{L_{j},k_{4}})^2}V(r_c),
\end{equation}
\begin{equation}\label{eq:SM16}
    \bar{N} = N_{n_i,\lvert \ell_i \rvert} N_{n_j,\lvert \ell_j \rvert} N_{N_i,\lvert L_i \rvert} N_{N_j,\lvert L_j\rvert}.
\end{equation}
In order to carry out the integration, it is necessary to represent the Jacobi coordinates in the correct channel by introducing the following transformation,
\begin{align}\label{eq:SM17}
    \mathbf{r}_{i} & = \gamma_{ij}\mathbf{r}_j + \sigma_{ij}\mathbf{R}_j, \\ 
    \mathbf{R}_{i} & = \alpha_{ij}\mathbf{r}_j + \beta_{ij}\mathbf{R}_j.
\end{align}
The coefficients can be extracted from the Jacobi Coordinates definitions, obtaining,

\begin{align*}
\gamma_{11} &= \gamma_{22} = \gamma_{33} = 
\beta_{11} = \beta_{22} = \beta_{33} = 1, \\
\sigma_{11} &= \sigma_{22} = \sigma_{33} = 
\alpha_{11} = \alpha_{22} = \alpha_{33} = 0, \\
\gamma_{12} &= -\frac{m_1}{m_1 + m_3}, & \sigma_{12} &= -1, \\
\gamma_{23} &= -\frac{m_2}{m_2 + m_1}, & \sigma_{23} &= -1, \\
\gamma_{31} &= -\frac{m_3}{m_3 + m_2}, & \sigma_{31} &= -1, \\
\gamma_{21} &= -\frac{m_2}{m_2 + m_3}, & \sigma_{21} &= +1, \\
\gamma_{13} &= -\frac{m_1}{m_1 + m_2}, & \sigma_{13} &= +1, \\
\gamma_{32} &= -\frac{m_3}{m_3 + m_1}, & \sigma_{32} &= +1,
\end{align*}

\begin{align*}
\alpha_{12} &= 1 - \gamma_{21}\gamma_{12}, & \beta_{12} &= \gamma_{21}, \\
\alpha_{23} &= 1 - \gamma_{32}\gamma_{23}, & \beta_{23} &= \gamma_{32}, \\
\alpha_{31} &= 1 - \gamma_{13}\gamma_{31}, & \beta_{31} &= \gamma_{13}, \\
\alpha_{21} &= -1 + \gamma_{12}\gamma_{21}, & \beta_{21} &= \gamma_{12}, \\
\alpha_{13} &= -1 + \gamma_{31}\gamma_{13}, & \beta_{13} &= \gamma_{31}, \\
\alpha_{32} &= -1 + \gamma_{23}\gamma_{32}, & \beta_{32} &= \gamma_{23}.
\end{align*}

Imposing the transformations onto the matrix element, integrating $\mathbf{R}_c$ and the angular contribution yields,
\begin{equation}\label{eq:SM18}
    I^{ij}_{c,V} = \frac{2\pi^2}{\zeta^{ij}_{c}}e^{\frac{P^2}{\zeta^{ij}_{c}}} \int d\mathbf{r}_c \, e^{-\eta^{\prime \, ij}_{c} r^2_c}V(r_c) I_{0}(2Qr_c).
\end{equation}
where $I_0(x)$ is the zeroth order modified Bessel function of the first kind and,
\begin{align}\label{eq:SM19}
\mathbf{P} &= \sum_{i=1}^{4} p_i \, \epsilon_i \, \mathbf{D}_i,  \\
\mathbf{Q} &= \sum_{i=1}^{4} q_i \, \epsilon_i \, \mathbf{D}_i,
\end{align}
with
\begin{align}\label{eq:SM20}
p_1 = \nu_{n_i} \sigma_{ic}, \quad 
p_2 = \lambda_{N_i} \beta_{ic}, \quad 
p_3 = \nu_{n_j} \sigma_{jc}, \quad 
p_4 = \lambda_{N_j} \beta_{jc}
\end{align}
\begin{align}\label{eq:SM21}
q_1 = \nu_{n_i} \gamma_{ic} - f p_1, \quad 
q_2 = \lambda_{N_i} \alpha_{ic} - f p_2, \quad 
q_3 = \nu_{n_j} \gamma_{jc} - f p_3, \quad 
q_4 = \lambda_{N_j} \alpha_{jc} - f p_4
\end{align}
where \( f = \xi^{ab}_c / \zeta^{ab}_c \) and other quantities are defined as,
\begin{align}
\eta^{ab}_{c} &= \nu_a \gamma_{ac}^2 + \lambda_a \alpha_{ac}^2 + \nu_b \gamma_{bc}^2 + \lambda_b \alpha_{bc}^2, \\
\zeta^{ab}_{c} &= \nu_a \sigma_{ac}^2 + \lambda_a \beta_{ac}^2 + \nu_b \sigma_{bc}^2 + \lambda_b \beta_{bc}^2, \\
\xi^{ab}_{c} &= \nu_a \gamma_{ac} \sigma_{ac} + \lambda_a \alpha_{ac} \beta_{ac} + \nu_b \gamma_{bc} \sigma_{bc} + \lambda_b \alpha_{bc} \beta_{bc},
\end{align}
with $\eta^{\prime \, ab}_{c} = \eta^{ab}_c - \xi^{ab}_c/\zeta^{ab}_c$.
Next, we expand the Bessel function in its polynomial representation and pick out the terms that obey the properties of the IGSL functions, namely, the upper bound of the polynomial expansion is $L_{sum} = |\ell_i|+|L_i|+|\ell_j|+|L_j|$ and write $I^{ij}_{c,V}$ as,
\begin{equation}
    I^{ij}_{c,V}=\frac{2\pi^2}{\zeta^{ij}_{c}}e^{\frac{P^2}{\zeta^{ij}_{c}}} \sum_{m=0}^{L_{sum}}\frac{Q^{2m}}{2^{2m+1}(m!)^2}V_{m}(\eta^{\prime\, ij}_{c})
\end{equation}
and,
\begin{equation}
    V_{m}(\eta^{\prime\, ij}_{c}) = \int dr\, r^{2m+1}V(r_c)e^{-\eta^{\prime \, ij}_c r^2}. 
\end{equation}
Next, in order to facilitate the avoidance of round-off error by analytically calculating the $\epsilon$'s limit (cf. equation~\eqref{eq:SM14}), we will apply the \textit{upon the shoulder} technique. 
\begin{gather}
    \sum_{m = 0}^{L_{sum}}V_{m}(\eta^{\prime\, ab}_{c}) Q^{2m} = \sum_{m = 0}^{L_{sum}}w_{m} e^{\mu_m Q^{2m}} = \sum_{m = 0}^{L_{sum}}w_{m} \sum_{m=0}^{L_{sum}}\frac{1}{m!}(\mu_m Q^{2m})^{m},
\end{gather}\label{uponshould}
we choose $\mu_n=\mu_0\,c^n$ with $\mu_0 = 0.08$ and $c = 1.6$, then, $w_n$ can be determined by solving,
\begin{gather}
    V_{m} = \sum_{n}\frac{\mu^{m}_n}{m!} w_{n}.
\end{gather}
Using this representation, we have,
\begin{equation}
     I^{ij}_{c,V} =\frac{2\pi^2}{\zeta^{ij}_{c}}\sum_{k = 0}^{L_{sum}}w_{k}\exp{\left(\sum_{m<n=1}^{4}\mathcal{G}_{mn,j} \epsilon_m \epsilon_n \mathbf{D}_{m}\cdot \mathbf{D}_{n}\right)},
\end{equation}
\begin{equation}\label{eqG}
    \mathcal{G}_{ij,n} = \frac{2p_i p_j}{\zeta^{ij}_c} + \frac{2\mu_n q_i q_j}{\eta^{\prime \, ij}_c}.
\end{equation}
Next, we can expand the exponential and retain only the term of order, 
\begin{equation}
    \exp{\left(\sum_{m<n=1}^{4}\mathcal{G}_{mn,j} \epsilon_m \epsilon_n \mathbf{D}_{n}\cdot \mathbf{D}_{m}\right)} \sim \frac{\left(\sum_{m<n=1}^{4}\mathcal{G}_{mn,j} \epsilon_m \epsilon_n \mathbf{D}_{n}\cdot \mathbf{D}_{m}\right)^{L_{sum}}}{L_{sum}!},
\end{equation}
and applying the multinomial expansion formula, with some extra algebra and summing the coefficients associated with the same powers of $g_{ij}$'s, we have the final expression for the matrix element, 
\begin{align}\label{eqpot}
   \mathcal{V}^{(ij)}_{n_i,N_i,\ell_i,L_i;n_j,N_j,\ell_j,L_j} =\frac{2\pi^2 \bar{N}}{\zeta^{ij}_{c}\nu^{\lvert \ell_i\rvert}_{n_i}\nu^{\lvert \ell_j\rvert}_{n_j}\lambda^{\lvert L_i\rvert}_{N_i}\lambda^{\lvert L_j\rvert}_{N_j}} 
\nonumber\\  \times\sum_{n=1}^{n_{max}}S^{J,n}_{\ell_{i},\ell_{j},L_{i},L_{j}}\sum_{k=0}^{L_{sum}}w_{k}\,\mathcal{G}^{a^{(n)}_{12}}_{12,k}\mathcal{G}^{a^{(n)}_{13}}_{13,k}\mathcal{G}^{a^{(n)}_{14}}_{14,k}\mathcal{G}^{a^{(n)}_{23}}_{23,k}\mathcal{G}^{a^{(n)}_{24}}_{24,k}\mathcal{G}^{a^{(n)}_{34}}_{34,k}.
\end{align}
where,
\begin{align}
S^{J,n}_{\ell_{i},\ell_{j},L_{i},L_{j}}
&=\sum_{k_a K_a}\sum_{k_b K_b}
C_{\ell_i,k_i}\,C_{\ell_j,k_j}\,
C_{L_i,K_i}\,C_{L_j,K_j}
\nonumber\\
&\times
\sum_{a_{12}}\sum_{a_{13}}\sum_{a_{14}}
\sum_{a_{23}}\sum_{a_{24}}\sum_{a_{34}}
\frac{1}{
a_{12}!\,a_{13}!\,a_{14}!\,
a_{23}!\,a_{24}!\,a_{34}!}
\nonumber\\
&\times
\bigl(\mathbf{D}_{1}\!\cdot\!\mathbf{D}_{2}\bigr)^{a_{12}}
\bigl(\mathbf{D}_{1}\!\cdot\!\mathbf{D}_{3}\bigr)^{a_{13}}
\bigl(\mathbf{D}_{1}\!\cdot\!\mathbf{D}_{4}\bigr)^{a_{14}}
\bigl(\mathbf{D}_{2}\!\cdot\!\mathbf{D}_{3}\bigr)^{a_{23}}
\bigl(\mathbf{D}_{2}\!\cdot\!\mathbf{D}_{4}\bigr)^{a_{24}}
\bigl(\mathbf{D}_{3}\!\cdot\!\mathbf{D}_{4}\bigr)^{a_{34}}
\nonumber\\
&\times
\delta_{a_{12}+a_{13}+a_{14},\,\lvert \ell_i \rvert}\,
\delta_{a_{12}+a_{23}+a_{24},\,\lvert L_i \rvert}\,
\delta_{a_{13}+a_{23}+a_{34},\,\lvert \ell_j \rvert}\,
\delta_{a_{14}+a_{24}+a_{34},\,\lvert L_j \rvert}
\nonumber\\
&\times
\delta_{a_{12},\,a_{12}^{(n)}}\,
\delta_{a_{13},\,a_{13}^{(n)}}\,
\delta_{a_{14},\,a_{14}^{(n)}}\,
\delta_{a_{23},\,a_{23}^{(n)}}\,
\delta_{a_{24},\,a_{24}^{(n)}}\,
\delta_{a_{34},\,a_{34}^{(n)}}.
\end{align}
are the set of coefficients which are generated by the multinomial expansion.
The re-framing of the matrix element in terms of combinatorial coefficients and shift parameters allows for precalculation of the $S$ term, which in turn yields a very efficient computational scheme. 
All other matrix elements are developed in an analogous way, and we will just present the final result for each. First, the overlap element,
\begin{align}
    \mathcal{N}^{(i j)}_{n_i,N_i,\ell_i,L_i;n_j,N_j,\ell_j,L_j} 
    = \frac{\pi^2 \bar{N}}{\zeta^{ij}_j\eta^{\prime\,ij}_j
        \nu^{\lvert \ell_i\rvert}_{n_i}
        \nu^{\lvert \ell_j\rvert}_{n_j}
        \lambda^{\lvert L_i\rvert}_{N_i}
        \lambda^{\lvert L_j\rvert}_{N_j}} \sum_{n=1}^{n_{\text{max}}}
    S^{J,n}_{\ell_{i},\ell_{j},L_{i},L_{j}}
    \mathcal{F}^{a^{(n)}_{12}}_{12}
    \mathcal{F}^{a^{(n)}_{13}}_{13}
    \mathcal{F}^{a^{(n)}_{14}}_{14}
    \mathcal{F}^{a^{(n)}_{23}}_{23}
    \mathcal{F}^{a^{(n)}_{24}}_{24}
    \mathcal{F}^{a^{(n)}_{34}}_{34},
\end{align}
where,
\begin{equation}
    \mathcal{F}_{ij} = \frac{2p_ip_j}{\zeta^{ij}_j}+\frac{2q_iq_j}{\eta^{\prime \, ij}_j}.
\end{equation}
And, for the kinetic energy,
\begin{align}
    \mathcal{K}^{(ij)}_{n_i,N_i,\ell_i,L_i;n_j,N_j,\ell_j,L_j} =\frac{\pi^2 \bar{N}}{\eta^{\prime \, ij}_{c}\zeta^{ij}_{c}\nu^{\lvert \ell_i\rvert}_{n_i}\nu^{\lvert \ell_j\rvert}_{n_j}\lambda^{\lvert L_i\rvert}_{N_i}\lambda^{\lvert L_j\rvert}_{N_j}}\sum_{n=1}^{n_{max}}S^{J,n}_{\ell_{i},\ell_{j},L_{i},L_{j}}\sum_{k=0}^{M_{sum}}w_{k}\,t^{a^{(n)}_{12}}_{12,k}t^{a^{(n)}_{13}}_{13,k}t^{a^{(n)}_{14}}_{14,k}t^{a^{(n)}_{23}}_{23,k}t^{a^{(n)}_{24}}_{24,k}t^{a^{(n)}_{34}}_{34,k},
\end{align}
where $w_k$ is determined using the \textit{upon the shoulder} method
\begin{equation}
   t_{ij,n} = \frac{2p_ip_j}{\zeta^{ij}_{j}}+\frac{2q_iq_j}{\eta^{\prime \, ij}_{j}}+\frac{\mu_n K_{ij}}{K_0},
\end{equation}
\begin{align}
K_{0} &=
  4\,\tau_{r_j}\,\nu_{j}\!\left(\frac{\nu_{j}}{\eta^{\prime \,ij}_j}-1\right)
  + 4\,\tau_{R_j}\,\lambda_{j}\!\left(\frac{\lambda_{j}}{\zeta^{\prime \,ij}_j}-1\right),
\\[6pt]
K_{mn} &=
  \frac{4\,\tau_{r_n}\,\nu_{j}^{2}}{\eta'}\!
  \Bigl(
      q_{m}q_{n}
      - q_{1}\delta_{m,1}\delta_{n,3}
      - q_{2}\delta_{m,2}\delta_{n,3}
      - q_{4}\delta_{m,3}\delta_{n,4}
  \Bigr)
\nonumber\\
&\quad+
  \frac{4\,\tau_{R_n}\,\lambda_{j}^{2}}{\zeta'}\!
  \Bigl(
      q'_{m}q'_{n}
      - q'_{1}\delta_{m,1}\delta_{n,3}
      - q'_{2}\delta_{m,2}\delta_{n,3}
      - q'_{4}\delta_{m,3}\delta_{n,4}
  \Bigr),
\end{align}
where, 
\begin{equation}
    \tau_{r_n} = \frac{\hbar^2}{2 \mu_{r_n}} \quad\text{and}\quad
    \tau_{R_n} = \frac{\hbar^2}{2 \mu_{R_n}}\,.
\end{equation}
and,
\begin{align*}
p_1 &= \nu_a \sigma_{ab}, & 
p_2 &= \lambda_a \beta_{ab}, & 
p_3 &= 0, & 
p_4 &= \lambda_b, \\
p_1' &= \nu_a \gamma_{ab}, & 
p_2' &= \lambda_a \alpha_{ab}, & 
p_3' &= \nu_b, & 
p_4' &= 0, \\
\mathbf{Q} &= \mathbf{P}' - (\xi^{ij}_{j}/\zeta^{ij}_{j})\mathbf{P}, & 
\mathbf{Q}' &= \mathbf{P}' - (\xi^{ij}_{j}/\eta^{ij}_{j})\mathbf{P}'.
\end{align*}
In the article, observables such as the root-mean-squared radii for some channel $c$, $\langle r^2_c\rangle$, are calculated in order to determine the geometry of the bound-state. The Gaussian Expansion Method makes such matrix elements simple to calculate. Here, we will generalize to any power of $r_c$, as the method yields analytical integrals, 

\begin{equation}
    \bra{\Psi_{J}} r^\lambda_c\ket{\Psi_{J}} = \sum_{a,\alpha}\sum_{\alpha^{\prime},b}\mathcal{A}^{a}_{\alpha}\mathcal{A}^{b}_{\alpha^{\prime}}\bra{\Phi^{a}_{\alpha}} r^2_c\ket{\Phi^{b}_{\alpha^{\prime}}},
\end{equation}
in analogy to Eq.~(\ref{eqpot}), we write, 
\begin{eqnarray}
    \bra{\Psi_{J}} r^\lambda_c\ket{\Psi_{J}} = \sum_{a,\alpha,\alpha^{\prime},b}\frac{2\pi^2 \bar{N}\mathcal{A}^{a}_{\alpha}\mathcal{A}^{b}_{\alpha^{\prime}}}{\zeta^{ab}_{c}\nu^{\lvert \ell_a\rvert}_{n_a}\nu^{\lvert \ell_b\rvert}_{n_b}\lambda^{\lvert L_a\rvert}_{N_a}\lambda^{\lvert L_b\rvert}_{N_b}} 
\sum_{n=1}^{n_{max}}S^{J,n}_{\ell_{a},\ell_{b},L_{a},L_{b}}
\nonumber\\
\times \sum_{k=0}^{M_{sum}}R^{c,\lambda}_{k}\,\mathcal{G}^{a^{(n)}_{12}}_{12,k}\mathcal{G}^{a^{(n)}_{13}}_{13,k}\mathcal{G}^{a^{(n)}_{14}}_{14,k}\mathcal{G}^{a^{(n)}_{23}}_{23,k}\mathcal{G}^{a^{(n)}_{24}}_{24,k}\mathcal{G}^{a^{(n)}_{34}}_{34,k}.
\end{eqnarray}
Continuing the analogy, $R^{c,\lambda}_k$ is determined by inverting the following equation,
\begin{equation}
    V^{c,\lambda}_{j} = \sum_{m}\frac{\mu^{j}_m}{j!} R^{c}_{m},
\end{equation}
with,
\begin{align}
    V^{c,\lambda}_{j} &= \int dr\, r^{2j+1}r^{\lambda}_ce^{-\eta^{\prime}_c r^2} = \frac{\Gamma\left(\frac{2(j+1)+\lambda}{2}\right)}{2 \left(\eta^{\prime}_c\right)^{2(j+1)+\lambda}},
\end{align}
and $\mathcal{G}_{ij,n}$ defined as in Eq.~\eqref{eqG}. 